\documentclass[review,3p,sort&compress]{elsarticle}
\biboptions{sort&compress}

\usepackage{graphicx,wrapfig,lipsum}
\usepackage{mathtools}
\usepackage{mathrsfs} 
\usepackage{amsmath,amssymb,bm,accents}
\usepackage{amsthm}
\usepackage[version=4]{mhchem}
\usepackage{chemformula}
\usepackage{siunitx}
\usepackage{xurl}
\usepackage{booktabs} 
\usepackage{longtable} 

\numberwithin{equation}{section}

\def\restrict#1{\raise-.5ex\hbox{\ensuremath|}_{#1}}

    \DeclareMathSymbol{\widetildesym}{\mathord}{largesymbols}{"65}
    \newcommand\lowerwidetildesym{%
      \text{\smash{\raisebox{-1.5ex}{%
        $\widetildesym$}}}}
            \newcommand\bigtilde[1]{%
              \mathchoice
                {\accentset{\displaystyle\lowerwidetildesym}{#1}}
        {\accentset{\textstyle\lowerwidetildesym}{#1}}
        {\accentset{\scriptstyle\lowerwidetildesym}{#1}}
        {\accentset{\scriptscriptstyle\lowerwidetildesym}{#1}}
    }
 
\begin{document}
\journal{AIChE Journal}
\title{Consolidated theory of fluid thermodiffusion}
\date{\today}
\author[maths,FI]{Alexander Van-Brunt}
\author[maths]{Patrick E. Farrell}
\author[engsci,FI]{Charles W. Monroe}
\ead{charles.monroe@eng.ox.ac.uk}
\address[maths]{Mathematical Institute, University of Oxford, Woodstock Road, Oxford, OX2 6GG, UK}
\address[engsci]{Department of Engineering Science, University of Oxford, Oxford, OX1 3PJ, United Kingdom}
\address[FI]{The Faraday Institution, Harwell Campus, Didcot, OX11 0RA, United Kingdom}

\begin{abstract}
We present the Onsager--Stefan--Maxwell thermodiffusion equations, which account for the Soret and Dufour effects in multicomponent fluids. Unlike transport laws derived from kinetic theory, this framework preserves the structure of the isothermal Stefan--Maxwell equations, separating the thermodynamic forces that drive diffusion from the force that drives heat flow. The Onsager--Stefan--Maxwell transport-coefficient matrix is symmetric, and the second law of thermodynamics imbues it with simple spectral characteristics. This new approach  allows for heat to be considered as a pseudo-species and  proves equivalent to both the intuitive extension of Fick's law and the generalized Stefan--Maxwell equations popularized by Bird, Stewart, and Lightfoot. A general inversion process facilitates the unique formulation of flux-explicit transport equations relative to any choice of convective reference velocity. Stefan--Maxwell diffusivities and thermal diffusion factors are tabulated for gaseous mixtures containing helium, argon, neon, krypton, and xenon. The framework is deployed to perform numerical simulations of steady three-dimensional thermodiffusion in a ternary gas.
 \end{abstract}

\maketitle


\section{Introduction}

Thermodiffusion generally describes situations where heat transfer is coupled to mass transfer, a class of phenomena first recognized by Ludwig \cite{ludwig1856diffusion}. Primarily it refers to the process by which a temperature gradient drives diffusion, known as the Soret effect  \cite{Soret1879}, and the converse process, by which a concentration gradient drives irreversible heat flux, called the Dufour effect \cite{Dufourpaper}. Rahman and Saghir provide a concise historical overview of thermodiffusion research over the last century \cite{Soreteffecthistory}. Interest continues today: recently Platten discussed a host of new experimental techniques for probing thermodiffusion \cite{Soretreview2005}; K\"ohler and Morozov reviewed measurements for organic solutions \cite{Soreteffectreview2016}.  

Chemical engineers have pursued separation processes based on thermodiffusion for decades \cite{Curtiss1999}. Chemical physicists, however, utilized the phenomenon first, exploiting the Soret effect for isotope separation \cite{Isotope1939}. Thermodiffusion finds contemporary applications in semiconductor fabrication \cite{Semiconductorreview1, Semiconductorreview2}. In electrochemistry, thermoelectric phenomena such as the Peltier effect have been tied to Soret and Dufour fluxes \cite{newman1995thermal}. Metallurgists have investigated thermodiffusion in liquid alloys \cite{ThermoDLiquidmetal1955}. Interest has also surged among biochemical researchers, who study thermophoresis of biomolecules \cite{Microscalethermophoresis2014}. 

Gas-phase thermodiffusion has a sound theoretical foundation, laid by Chapman and Enskog's kinetic theory \cite{ChapmanCowling}. Hirschfelder, Curtiss, and Bird implemented the kinetic theory for multicomponent gases, obtaining transport laws that account for coupled heat flux and mass flux \cite{hirschfelder1954molecular}. These and an alternative, flux-explicit formulation are covered in the textbook of Bird, Stewart, and Lightfoot \cite{BSL2nd}. A broader framework for coupled heat and mass transport was created by Onsager \cite{Onsager1931second}, whose theory of irreversible thermodynamics applies without modification to nonideal and condensed phases, as well as perfect gases. 

Structural differences among the available theoretical frameworks pose a barrier to experimentalists, whose work has been constrained almost entirely to thermodiffusion in binary systems. Importantly, properties measured in one embodiment of the theory do not readily translate into another. The numerical complexity of multicomponent simulations has also impeded progress.  

Here we set out to unify the major continuum-scale theories of thermodiffusion, and to deploy the elementary principles of irreversible thermodynamics to produce more accurate and robust numerical methods. As a practical matter, the unification effort reveals how to interconvert property sets measured under different paradigms. 

We propose a new constitutive scheme---the Onsager--Stefan--Maxwell thermodiffusion equations---that simplifies analysis and numerics by accounting for heat as a pseudo-species. This provides a natural route by which binary-system properties can be extrapolated to simulate multicomponent systems. For 10 noble-gas pairs, we tabulate the properties needed to simulate binary thermodiffusion with any of the popular theories and in addition detail how to extrapolate these properties to simulate $n$-ary systems. We demonstrate the applicability of the Onsager--Stefan--Maxwell framework by simulating steady, three-dimensional thermodiffusion in a ternary noble-gas mixture.

\section{Approaches to thermodiffusion}

The typical routes taken when creating thermodiffusion models are: derivation by methods of irreversible thermodynamics; derivation from kinetic theory; and intuitive formulation by extending Fick's law and Fourier's law. Each of these choices may have particular benefits and disadvantages with regard to physical rigour, ease of parametrization, and computational simplicity. Our task of reconciling these different perspectives begins with summarizing the available theories.

\subsection{Irreversible thermodynamics}
Transport laws based on irreversible thermodynamics derive from the local entropy balance. Following the derivation of Jaumann \cite{Jaumann1911}, one begins with the entropy continuity equation
\begin{equation}
\frac{\partial \rho \hat{S}}{\partial t} = - \vec{\nabla} \cdot \vec{N}_{S} + \dot{s},
\end{equation}
where $\rho$ is the local mass density of the phase, $\hat{S}$ the specific entropy,  $\vec{N}_{S}$ the total entropy flux, and $\dot{s}$ the volumetric rate of entropy production. Through a lengthy thermodynamic analysis involving the first law, the volumetric equation of state, Euler's extensivity theorem, and the Gibbs--Duhem equation, as well as various Legendre transformations and Maxwell relations \cite{hirschfelder1954molecular,Goyal01012017}, entropy continuity transforms into an expression that describes the instantaneous energy dissipation. For an isothermal viscous-fluid diffusion system comprising $n$ chemical species under a null rate of deformation, the dissipation rate $T \dot{s}$ has units of power per volume, $\si{J.m^{-3}.s^{-1}}$, and takes the form
\begin{equation} \label{Fundmentalentropyproduction}
T \dot{s} = -\vec{\nabla} \ln T \cdot \vec{q}\hspace{1.5pt}' +  \sum_{i=1}^{n} \vec{d}_{i} \cdot \vec{v}_{i},
\end{equation} 
in which $T$ is the absolute temperature. Here $\vec{d}_{i}$ denotes the thermodynamic force that drives diffusion of species $i$, with units $\si{N.m^{-3}}$ (or $\si{J.m^{-4}}$), and $\vec{v}_{i}$ is its species velocity  (with respect to a reference velocity yet to be defined);  $\vec{q}\hspace{1.5pt}'$ denotes the irreversible heat flux, with units $\si{\joule.m^{-2}.s^{-1}}$. 

To ensure consistency of units among the fluxes in the dissipation functional, we define the \emph{thermal velocity} $\vec{v}_0$ as
\begin{equation} \label{thermalv}
\vec{v}_{0} = \frac{\vec{q}\hspace{1.5pt}'}{\rho \hat{C}_{p} T},
\end{equation} 
where $\hat{C}_{p}$ is the phase's specific constant-pressure heat capacity, in $\si{J.kg^{-1}.K^{-1}}$ (or $\si{m^2.s^{-2}.K^{-1}}$). This necessitates that the conjugate \emph{thermal driving force $\vec{d}_0$} be 
\begin{equation} \label{d0ofT}
\vec{d}_{0} =- \rho \hat{C}_{p} \vec{\nabla} T,
\end{equation}
simplified with the identity $T \vec{\nabla} \ln T = \vec{\nabla} T$. By construction, $\vec{v}_{0}$ and $\vec{d}_{0}$ have units that match the quantities $\vec{v}_{i}$ and $\vec{d}_{i}$ defined earlier. Henceforth we call members of the collection $\{ \vec{d}_{i} \}_{i=0}^{n}$ \textit{thermodiffusion driving forces}, and those of $\{ \vec{v}_{i} \}_{i=0}^{n}$ \textit{thermodiffusive velocities}; we reserve the terms \emph{diffusion driving forces} and \emph{species velocities} to refer to the smaller collections $\{ \vec{d}_{i} \}_{i=1}^{n}$ and $\{ \vec{v}_{i} \}_{i=1}^{n}$, respectively. A phase containing $n$ species affords $n+1$ thermodiffusive driving forces and $n+1$ thermodiffusive velocities.

The principles of irreversible thermodynamics require that proper constitutive laws for transport be formulated with the conjugate pairs of fluxes and forces that appear in the dissipation functional \cite{deGrootMazur}. In equation \eqref{Fundmentalentropyproduction}, for instance, the thermodiffusive velocity $\vec{v}_{i}$ is conjugate to the respective thermodiffusion driving force $\vec{d}_{i}$. Onsager postulated that, within nonuniform systems not too far from equilibrium, global thermodynamic state functions can be applied to describe the local states \cite{Onsager1931second,Onsager1945}. Under this condition of `microscopic equilibrium', Onsager posited further that a linear transformation maps the thermodiffusive velocities into their conjugate driving forces. 
 
More concretely, thermodiffusion constitutive laws from irreversible thermodynamics take the force-explicit forms 
\begin{eqnarray} 
\label{Forceexplicit2} 
\vec{d}_{0} = \bigtilde{M}_{00} \vec{v}_{0} + \sum\limits_{j=1}^n \bigtilde{M}_{0j} \vec{v}_{j} \\ 
\label{Forceexplicit1}
\vec{d}_{i} = \bigtilde{M}_{i0} \vec{v}_{0} + \sum\limits_{j=1}^n \bigtilde{M}_{ij} \vec{v}_{j} ,
\end{eqnarray}
 where the second equation holds for each $i \in \left\{ 1, ..., n \right\}$.\footnote{For simplicity we have assumed when writing equations \eqref{Forceexplicit2} and \eqref{Forceexplicit1} that the material being modelled is isotropic, in which case the coefficients $\bigtilde{M}_{ij}$ are scalars. In general, anisotropic systems can be handled by replacing every $\bigtilde{M}_{ij}$ with a symmetric tensor \cite{Onsager1931first}, and replacing all scalar multiplications on the right with tensor-vector products.} 
These relationships,  which we name the \emph{Onsager--Stefan--Maxwell thermodiffusion equations}, introduce the Onsager drag coefficients $\bigtilde{M}_{ij}$, a set of material properties with units $\si{J.s.m^{-5}}$ (i.e., force per volume per velocity) that parametrizes how the thermodiffusive velocities map into the thermodiffusive driving forces. These equations generalize the isothermal Onsager--Stefan--Maxwell equations
\begin{equation}
\vec{d}_{i} = \sum\limits_{j=1}^n {M}_{ij} \vec{v}_{j},
\end{equation}
for $i \in \{1, \dots, n\}$, expressed in terms of the isothermal drag coefficients $M_{ij}$~\cite{Lightfoot1962}.
Note that equations \eqref{Forceexplicit2} and \eqref{Forceexplicit1} have not been written as compactly as possible because in later discussion it will sometimes be convenient to consider the equation for the thermal force $\vec{d}_0$ separately, and to keep the coefficients whose second index equals $0$ out of the sums.

It is often useful to think of the set of all Onsager drag coefficients as being arrayed within an $(n+1) \times (n+1)$ matrix $\bigtilde{\textbf{M}}$ whose row and column indices range across the integers from $0$ to $n$, which we will call the \emph{non-isothermal drag matrix}. Onsager asserts that with a proper choice of velocities and driving forces based in the energy dissipation, such as equation \eqref{Fundmentalentropyproduction}, the reciprocal relation among the transport coefficients is expressed by symmetry of $\bigtilde{\textbf{M}}$ \cite{Onsager1931second}. This symmetry has practical implications for the spectral structure of the thermodiffusion problem.

Onsager's formulation is not as abstract as it first appears. Hirschfelder et al.~substantiated the general concepts, manipulating thermodynamic laws and hydrodynamic equations to cast the dissipation functional in terms of standard quantities from thermodynamics \cite{hirschfelder1954molecular,Goyal01012017}. The general diffusion driving forces within a non-isobaric, non-isothermal, multicomponent fluid are  \cite{hirschfelder1954molecular,Goyal01012017}
\begin{equation}\label{didef}
\vec{d}_{i} =   -c_{i} \left( \vec{\nabla} \mu_{i} + \overline{S}_{i} \vec{\nabla} T - \frac{\overline{m}_{i}}{\rho} \vec{\nabla} p \right),
\end{equation}
where the index $i$ ranges from 1 to $n$. In this definition, $p$ denotes the external pressure; for each species $i$, $c_{i}$ is the molar concentration, $\overline{m}_{i}$ the molar mass, $\mu_{i}$ the chemical potential, and $\overline{S}_{i}$ the partial molar entropy. Thus practical implementation of laws \eqref{Forceexplicit2} and \eqref{Forceexplicit1} is possible if the thermodynamic state variables are known functions of $\left\{ c_i \right\}_{i=1}^n$, $T$, and $p$. Determining these state variables is sufficiently involved, however, that it is uncommon for experimentalists to perform enough measurements to specify the drag matrix $\bigtilde{\textbf{M}}$ for condensed phases. 

\subsection{The kinetic theory of gases}
Chapman--Enskog theory \cite{ChapmanCowling} leads to a set of thermodiffusion transport laws whose structure differs from laws \eqref{Forceexplicit2}-\eqref{Forceexplicit1}. Through the Chapman--Enskog perturbation of the Boltzmann equation for a multicomponent gas, Hirschfelder et al.~\cite{hirschfelder1954molecular} derived the so-called \emph{generalized Stefan--Maxwell equations}, which they wrote as 
\begin{equation} \label{GeneralMSequations}
\vec{d}_{i} - \sum_{j=1}^{n} \frac{RT c_{i} c_{j}}{\mathscr{D}_{ij} c_{\text{T}}} \left( \frac{D_{i}^{T}}{\rho_{i}}-\frac{D_{j}^{T}}{\rho_{j}} \right) \vec{\nabla} \ln T 
= \sum_{j=1}^{n} \frac{RT c_{i} c_{j}}{\mathscr{D}_{ij} c_{\text{T}}} \left( \vec{v}_{i} -\vec{v}_{j} \right)
\end{equation}
for $i \in \left\{ 1, ..., n \right\}$, where $R$ is the gas constant. Here $c_\textrm{T} = \sum_{i=1}^n c_i$ is the total molar concentration (inverse molar volume) and $\rho_{i}$ denotes the mass density of species $i$; $\rho_i = c_i \overline{m}_i$. Equation \eqref{GeneralMSequations} introduces another set of transport properties, namely, the Stefan--Maxwell diffusion coefficients $\mathscr{D}_{ij}$ and the coefficients of thermal diffusion $D_{i}^{T}$. 

As pointed out by Newman \cite{newman1995thermal}, the parameters $D_{i}^{T}$ have units of viscosity (\emph{i.e.} $\si{Pa.s}$, or $\si{J.s.m^{-3}}$, or $\si{kg.m^{-1}.s^{-1}}$), which is unconventional for a diffusion coefficient. To retain dimensional consistency and simplify notation in the sequel, we therefore define
\begin{equation} \label{Soretdiffusivitydef}
\mathscr{D}_i^T = \frac{D_i^T}{\rho_i},
\end{equation}
which we call the \emph{Soret diffusivity} of species $i$. The Soret diffusivity $\mathscr{D}_i^T$ has typical units, $\si{m^2.s^{-1}}$.

Some attempts have been made to bring constitutive formulation \eqref{GeneralMSequations} into harmony with \eqref{Forceexplicit2}-\eqref{Forceexplicit1}.  For isothermal multicomponent diffusion, Lightfoot et al.~\cite{Lightfoot1962} pointed out that the Stefan--Maxwell equations can be understood in terms of Onsager's transport equations by identifying the entries $M_{ij}$ of the isothermal transport matrix $\textbf{M}$ as
\begin{equation} \label{transport matrix}
M_{ij} = \begin{cases} -\frac{RT c_{i} c_{j}}{\mathscr{D}_{ij} c_{\text{T}}}  \:\:\: \text{if} \:\: i \neq j \\
\sum_{k \neq i}^{n}\frac{RT c_{i} c_{k}}{\mathscr{D}_{ik} c_{\text{T}}} \:\:\: \text{if} \:\: i = j. \end{cases}
\end{equation}
Onsager reciprocity therefore implies symmetry of the coefficients $\mathscr{D}_{ij}$. 

Identity \eqref{transport matrix} fails to carry over to the non-isothermal case because the generalized Stefan--Maxwell equations \eqref{GeneralMSequations} mix driving forces and fluxes, which may also violate the conditions required to ensure a symmetric Onsager reciprocal relation \cite{Coleman1960,MONROE20094804}. Specifically, equation \eqref{GeneralMSequations} contains no thermal velocity on the right; on the left, it also mixes the thermal driving force with a diffusion driving force. Therefore $\bigtilde{\textbf{M}} \neq \textbf{M}$, even if comparison is restricted to the submatrix with $ 1 \leq i,j \leq n$, as we shall see in section \ref{sec:properties}.

\subsection{Multicomponent Fick's law}
Many transport analyses begin in a more \emph{ad hoc} fashion, by postulating flux-explicit forms of the transport laws with the structure of Fick's law \cite{BSL2nd, taylor1993multicomponent, deGrootMazur}. For a general multicomponent mixture this results in expressions for the excess molar flux of species $i$, $\vec{J}_i$, of the form
\begin{equation} \label{Fickianansatz}
\vec{J}_i = - \bigtilde{D}_{i0} \vec{\nabla} T - \sum_{j=1}^{n} \bigtilde{D}_{ij} \vec{\nabla} c_{j} 
\end{equation}
for $i \in \left\{ 1, ..., n \right\}$. To close this thermodiffusion model, Fickian laws from equation \eqref{Fickianansatz} must be augmented by a complementary extension of Fourier's law of heat conduction, written as
\begin{equation} \label{Fourieransatz}
\vec{q}\hspace{1.5pt}' = - \bigtilde{D}_{00} \vec{\nabla} T - \sum_{j=1}^{n} \bigtilde{D}_{0j} \vec{\nabla} c_{j} ,
\end{equation}
which expresses the irreversible heat flux.

Clearly the entries of the $( n + 1) \times ( n + 1 )$ Fickian thermodiffusion matrix $\bigtilde{\textbf{D}}$ have inconsistent units. In the multicomponent generalization of Fick's law, equation \eqref{Fickianansatz}, the transport coefficients $\bigtilde{D}_{ij}$ that accompany concentration gradients can be seen as traditional diffusivities, in $\si{m^2.s^{-1}}$; the coefficients of thermal diffusion $\bigtilde{D}_{i0}$ have units $\si{mol.m^{-2}.s^{-1}.K^{-1}}$. In the generalization of Fourier's law from equation \eqref{Fourieransatz}, $\bigtilde{D}_{00}$ is an effective thermal conductivity, in $\si{J.m^{-1}.s^{-1}.K^{-1}}$, but the other $\bigtilde{D}_{0j}$ are Dufour coefficients, in $\si{J.m^2.s^{-1}.{mol}^{-1}}$. These discrepancies emphasize that the reciprocal relation here is ambiguous; it cannot simply be a symmetry of the $\bigtilde{\textbf{D}}$ matrix.

A few paths have been proposed to connect the Onsager--Stefan--Maxwell equations \eqref{Forceexplicit2}-\eqref{Forceexplicit1} with relations in the Fickian forms of equations \eqref{Fickianansatz}-\eqref{Fourieransatz}. Despite their linearity, transformation between force-explicit and flux-explicit constitutive formulations requires more than a simple inversion of the transport matrix. As pointed out by Helfand \cite{Helfand1960}, inversion is thwarted by the fact that a Gibbs--Duhem relation constrains the diffusion driving forces. When $\vec{d}_i$ is defined with equation \eqref{didef}, the Gibbs--Duhem equation requires that
\begin{equation} \label{Gibbs-Duhem}
\sum_{i=1}^{n} \vec{d}_{i} = 0, 
\end{equation}
which further implies singularity of the isothermal drag matrix $\textbf{M}$. A pseudo-inversion process is enabled, however, by considering kinematic constraints implied by the invariance of diffusion phenomena with respect to the velocity that drives convection.

Flux-law inversions implicitly based on the mass-average velocity were addressed by Curtiss and Bird \cite{Curtiss1999} and Fong et al.~\cite{Fong2020}. Inversions with respect to other reference velocities have been implemented by Helfand \cite{Helfand1960} and Newman \cite{newman2012electrochemical}. Even when applied in isothermal cases, however, these derivation procedures are somewhat impromptu. The Curtiss--Bird approach to inversion, which produces the most well known flux-explicit model for thermodiffusion, was only taken up to quaternary mixtures. No general method so far has been put forward to derive a flux-explicit form of the Onsager--Stefan--Maxwell equations relative to an arbitrary convective velocity.

\section{Properties of the non-isothermal drag matrix} \label{sec:properties}
First we set out to reconcile the generalized Stefan--Maxwell equations \eqref{GeneralMSequations} with the Onsager--Stefan--Maxwell thermodiffusion equations \eqref{Forceexplicit2}-\eqref{Forceexplicit1}. We then discuss how the second law of thermodynamics and the principle of convection invariance impose further structure beyond simple symmetry of the non-isothermal drag matrix $\bigtilde{\textbf{M}}$.

\subsection{Construction and connection to kinetic theory}

One can move toward the form of the generalized Stefan--Maxwell equations \eqref{GeneralMSequations} by isolating $\vec{v}_0$ from equation \eqref{Forceexplicit2} and substituting the result into equation \eqref{Forceexplicit1}. Regrouping terms and rearranging yields
\begin{equation} \label{Halfwayconstruct}
\vec{d}_{i} + \frac{\bigtilde{M}_{0i} \rho \hat{C}_{p} T}{\bigtilde{M}_{00}} \vec{\nabla} \ln T = \sum_{j=1}^{n} \left( \bigtilde{M}_{ij} -  \frac{\bigtilde{M}_{0i} \bigtilde{M}_{0j}}{\bigtilde{M}_{00}} \right) \vec{v}_{j} .
\end{equation}
Agreement of the left side with \eqref{GeneralMSequations} requires that
\begin{equation} \label{Startingtodefinethermalcolumn}
\frac{\bigtilde{M}_{0i} \rho \hat{C}_p T}{\bigtilde{M}_{00}} = - \sum_{j=1}^{n} \frac{RT c_{i} c_{j}}{ c_{\text{T}}\mathscr{D}_{ij}} \left( \mathscr{D}_i^T - \mathscr{D}_j^T \right).
\end{equation}
Insertion of equation \eqref{transport matrix}, followed by the simplification $\sum_{j=1}^n M_{ij} = 0$ implied by Gibbs--Duhem equation \eqref{Gibbs-Duhem} and the Onsager reciprocal relation, produces
\begin{equation} \label{CouplednonisothermalM}
\bigtilde{M}_{0i} = - \frac{\bigtilde{M}_{00}}{\rho \hat{C}_p T} \sum_{j=1}^{n} M_{ij} \mathscr{D}_{j}^T. 
\end{equation}
Similarly, agreement of the right side of equation \eqref{Halfwayconstruct} with \eqref{GeneralMSequations} demands for all $1 \leq i,j \leq n$ that
\begin{equation} \label{DiffusionnonisothermM}
\bigtilde{M}_{ij} = M_{ij} + \frac{\bigtilde{M}_{0i} \bigtilde{M}_{0j}}{\bigtilde{M}_{00}},
\end{equation}
where $M_{ij}$ relates to Stefan--Maxwell diffusivities through equation \eqref{transport matrix}. These conclusions can be summarized by writing the non-isothermal drag matrix in block form, as
\begin{equation} \label{Blocknonisotherm}
\bigtilde{\textbf{M}} =  \begin{bmatrix}
  \bigtilde{M}_{00}  & \widetilde{\textbf{m}}_{0}^{\top} \\
\widetilde{\textbf{m}}_{0} & \textbf{M} + \dfrac{\widetilde{\textbf{m}}_{0}\widetilde{\textbf{m}}_{0}^{\top}}{\bigtilde{M}_{00}}
\end{bmatrix},
\end{equation}
in which the column matrix $\widetilde{\textbf{m}}_{0} = \left[ \bigtilde{M}_{0i} \right]^{i=1}_n$ contains entries $\bigtilde{M}_{0i}$ defined by equation \eqref{CouplednonisothermalM}, superscript $\top$ indicates the matrix transpose,  and the first row's entries follow from the reciprocal relation $\bigtilde{\textbf{M}} = \bigtilde{\textbf{M}}^\top$.
This construction affirms that the generalized Stefan--Maxwell equations \eqref{GeneralMSequations} derived from kinetic theory are compatible with the Onsager--Stefan--Maxwell thermodiffusion laws \eqref{Forceexplicit2}-\eqref{Forceexplicit1} from irreversible thermodynamics.

 Despite the mixing of fluxes with driving forces in the generalized Stefan--Maxwell equations, one can show that their structure is consistent with the reciprocal relation $\bigtilde{\textbf{M}} = \bigtilde{\textbf{M}}^\top$. Substituting matrix \eqref{Startingtodefinethermalcolumn} into equation \eqref{Blocknonisotherm}, using flux law \eqref{Forceexplicit2}, then returning to the original thermal flux and driving force with definitions \eqref{thermalv}-\eqref{d0ofT}, yields
\begin{equation}
\vec{q}\hspace{1.5pt}' = -\frac{\rho^{2} \hat{C}_{\rho}^{2} T}{\bigtilde{M}_{00}} \nabla T - \sum_{i,j=1}^{n} \frac{RT c_{i} c_{j}}{ c_{\text{T}}\mathscr{D}_{ij}} \left( \mathscr{D}_i^T - \mathscr{D}_j^T \right) \vec{v}_{j}.
\end{equation}
This precisely matches the expression for irreversible heat flux provided by Hirschfelder, Curtiss, and Bird \cite{hirschfelder1954molecular}, although they define the coefficient of $-\vec{\nabla} T$ as the thermal conductivity. For now we retain the symbol $\bigtilde{M}_{00}$, leaving a detailed discussion of heat conduction to section \ref{subsec:thermalconductivity}.

\subsection{Entropy generation and spectral structure.}
A benefit of casting the generalized Stefan--Maxwell equations into the Onsager--Stefan--Maxwell form is the structure inherited by the transport matrix $\bigtilde{\textbf{M}}$. 
Given constitutive laws \eqref{Forceexplicit2} and \eqref{Forceexplicit1}, the entropy production due to thermodiffusion is equivalent to
\begin{equation} \label{dissipationMbar}
T \dot{s} = \sum_{i,j =0}^{n} \vec{v_{i}} \cdot \bigtilde{M}_{ij} \vec{v}_{j}.  
\end{equation} 
The second law of thermodynamics requires that $\dot{s} \geq 0$ for every collection of thermodiffusive velocities, so the matrix $\bigtilde{\textbf{M}}$ must be positive semidefinite, as well as symmetric. 

Direct calculation shows a connection between the spectral structures of the isothermal and non-isothermal drag matrices, $\textbf{M}$ and $\bigtilde{\textbf{M}}$, respectively. Insertion of matrix \eqref{Blocknonisotherm} into equation \eqref{dissipationMbar}, algebraic rearrangement of terms involving the symmetry of $\bigtilde{\textbf{M}}$,  and elimination of $\vec{v}_0$ with equation \eqref{Forceexplicit2} show that
\begin{equation} \label{dissipationMbar2}
T \dot{s} = \frac{1}{\bigtilde{M}_{00}} \big\| \vec{d}_0 \big\|^{2} + \sum_{i,j=1}^{n} \vec{v}_{i} \cdot M_{ij} \vec{v}_{j},
\end{equation}
in which $\|\vec{u}\|^{2}$ = $\vec{u} \cdot \vec{u}$ for any vector $\vec{u}$. Positive semidefiniteness of the non-isothermal matrix $\bigtilde{\textbf{M}}$ is thus implied by positive semidefiniteness of isothermal matrix $\textbf{M}$,  so long as the Onsager drag $\bigtilde{M}_{00}$ is positive. It is noteworthy that the sum on the right of equation \eqref{dissipationMbar2} can be expressed in terms of Stefan--Maxwell coefficients as  \cite{Vanbruntandother}
\begin{equation}\label{diffusionaldissipationMandD}
\sum_{i,j=1}^{n} \vec{v}_{i} \cdot M_{ij} \vec{v}_{j} = \frac{RT}{2c_\text{T}} \sum_{i,j=1}^{n} \frac{ c_{i} c_{j}  }{\mathscr{D}_{ij} } \left\| \vec{v}_{i} - \vec{v}_{j} \right\|^{2}.
\end{equation}
Thus, if all of the Stefan--Maxwell diffusivities in a material are constant, they must also all be positive. One should be wary that for materials where $\mathscr{D}_{ij}$ varies with composition, negative values are possible, and have been observed.

The distinction between convection and diffusion requires that no dissipation occurs unless there is relative motion of the species \cite{Onsager1945}. This feature naturally arises from the generalized Stefan--Maxwell equations, as may be seen from equation \eqref{diffusionaldissipationMandD} by noting that $\vec{v}_{i} - \vec{v}_{j} = \left(\vec{v}_{i} + \vec{u} \right) - \left(\vec{v}_{j} + \vec{u} \right)$ for any vector $\vec{u}$. In the isothermal case, invariance of the dissipation with respect to convection can equivalently be stated as a condition that $\textbf{M}$ has a single null eigenvalue corresponding to the eigenvector $ \textbf{1} = [1,1,...,1]^\top \in \mathbb{R}^{n}$, as noted by Helfand \cite{Helfand1960}. This spectral structure is also necessary to enforce Gibbs--Duhem equation \eqref{Gibbs-Duhem}. Indeed, in the isothermal case, equation \eqref{Forceexplicit1} reduces to
\begin{equation} \label{di_isothermal}
\vec{d}_{i}= \sum_{j=1}^{n} M_{ij} \vec{v}_{j}
\end{equation}
 for $i \in \left\{ 1, ..., n \right\}$,  and the facts that $\textbf{M} = \textbf{M}^\top$ and $\textbf{1}$ is a null eigenvector of $\textbf{M}$ imply equality \eqref{Gibbs-Duhem}. 

A similar conclusion about spectral structure can be drawn for the non-isothermal drag matrix. One can use equation \eqref{CouplednonisothermalM} to show that the sum over rows $1$ through $n$ of the first ($j=0$) column of $\bigtilde{\textbf{M}}$ satisfies
\begin{equation} 
\sum_{i=1}^{n} \bigtilde{M}_{i0} = \sum_{i,j=1}^{n} M_{ij} \mathscr{D}_{j}^T= 0.
\end{equation} 
Put another way, this says the column matrix $\textbf{1}$ is orthogonal to the previously defined column matrix $\widetilde{\textbf{m}}_{0}$, so that $\textbf{1}^\top \widetilde{\textbf{m}}_0 = 0$. Consequently, the block matrix from equation \eqref{Blocknonisotherm} must satisfy the matrix equation
\begin{equation} 
 \begin{bmatrix}
  \bigtilde{M}_{00}  & \widetilde{\textbf{m}}_{0}^\top \\
\widetilde{\textbf{m}}_{0} & \textbf{M} + \dfrac{\widetilde{\textbf{m}}_{0} \widetilde{\textbf{m}}_{0}^\top}{\bigtilde{M}_{00}}
\end{bmatrix}\begin{bmatrix}
0 \\ \textbf{1} 
\end{bmatrix} =  0 \cdot \begin{bmatrix}
0 \\ \textbf{1} 
\end{bmatrix},
\end{equation}
demonstrating that the the column matrix $[\begin{array}{cc} 0 & \textbf{1}^\top \end{array} ]^{\top}$ is a null eigenvector of the non-isothermal drag matrix. Like the isothermal case, this spectral structure of $\bigtilde{\textbf{M}}$ ensures the enforcement of Gibbs--Duhem relation \eqref{Gibbs-Duhem}. 

\section{Inverting the constitutive laws}

Many practitioners find it desirable to implement a flux-explicit (or velocity-explicit) transport formulation. As mentioned before, the transformation of force-explicit laws to a flux-explicit form is not entirely straightforward: since the isothermal drag matrix satisfies $\textbf{M} \textbf{1} = 0\textbf{1} = \textbf{o}$, where $\textbf{o}$ indicates a column of zeroes, it is singular, and consequently equation \eqref{di_isothermal} cannot be inverted directly. To resolve this issue it is necessary to consider how the flux-explicit laws behave with regard to the reference velocity selected to describe convection. Once the process of choosing a reference velocity is formalized, a kinematic relation derived from that choice can be exploited to develop a unique inverted formulation of the force-explicit transport laws. 

\subsection{Convective reference velocities} 
Convection is fundamentally anchored in species fluxes: a valid choice of convective velocity must always be a linear combination of the species velocities. Newman made the simplest choice, employing the $k^{\textrm{th}}$ species' velocity $\vec{v}_k$ as a reference for convection \cite{newman2012electrochemical}. Other common reference velocities include the mole-average velocity ${\vec{v}}\hspace{1pt}^c$ employed by Waldmann and Van Der Valk \cite{Waldmann1958, VANDERVALK1963417},
\begin{equation} \label{vcdef}
\vec{v}\hspace{1pt}^{c}  = \frac{1 }{c_\textrm{T}} \sum_{j=1}^{n} c_{j} \vec{v}_{j},
\end{equation}
 and the mass-average velocity ${\vec{v}}\hspace{1pt}^\rho$, defined as
\begin{equation} \label{vrhodef}
\vec{v}\hspace{1pt}^{\rho}  = \frac{1}{\rho} \sum_{j=1}^{n} \rho_{j} \vec{v}_{j},
\end{equation}
which was used by both Hirschfelder et al.~\cite{hirschfelder1954molecular} and Curtiss and Bird \cite{Curtiss1999}.

In Newman's inversion process \cite{newman2012electrochemical}, after species $k$ is selected as the reference for convection, it is shown that the isothermal drag matrix $\textbf{M}$ can be truncated by deleting its $k^{\textrm{th}}$ row and column without loss of information. This first minor matrix is nonsingular and can be inverted directly. Helfand considered isothermal, isobaric multicomponent diffusion \cite{Helfand1960}; when writing the Onsager--Stefan--Maxwell laws, he transformed the driving forces and selected a convective velocity to ensure that the null eigenvector of the flux-explicit transport-coefficient matrix coincided with that of the force-explicit matrix, enabling a Moore--Penrose pseudo-inversion.  Hirschfelder et al.~\cite{hirschfelder1954molecular} and Curtiss and Bird \cite{Curtiss1999} extensively manipulated the generalized Stefan--Maxwell equations to derive velocity-explicit thermodiffusion laws; the rather convoluted process they undertake is summarized in Bird, Stewart, and Lightfoot's book \cite{BSL2nd}. More recently Bothe et al.~\cite{Bothe2020OnTS} compared the isothermal Stefan--Maxwell equations with a generalized form of Fick's law and showed equivalent properties such as positivity preservation between the two approaches.

Depending on the context, different reference velocities may be desirable in flux-explicit constitutive laws \cite{BSL2nd,Cussler3rd}. Here we write a general reference velocity as $\vec{v}\hspace{1pt}^{\psi}$, following Goyal and Monroe \cite{Goyal01012017}. To create a convective velocity, one selects a collection of thermodynamic extensive quantities per unit volume $\left\{ \psi_i \right\}_{i=1}^n$, whose total volumetric amount $\psi_\textrm{T}$ is defined as
\begin{equation} \label{psiTdef}
\psi_{\textrm{T}} = \sum_{i=1}^{n} \psi_{i} = \textbf{1}^\top \boldsymbol{\psi},
\end{equation}
in which the rightmost expression introduces the column matrix $\boldsymbol{\psi} = \left[ \psi_i \right]^{i=1}_n$. Note that the quantities $\psi_{i}$ are not limited to choices with these precise physical interpretations; the only necessary restrictions on $\boldsymbol{\psi}$ are that it is real and that $\psi_\textrm{T} \neq 0$ to avoid degeneracy.

The $\psi$-average velocity, $\vec{v}\hspace{1pt}^{\psi}$, is defined as
\begin{equation} \label{vpsidef}
\vec{v}\hspace{1pt}^{\psi} =\frac{1}{\psi_{\textrm{T}}}  \sum_{j=1}^{n} \psi_{j} \vec{v}_{j}.
\end{equation} 
To further define the excess velocity of species $i$ relative to the $\psi$-average velocity, $\vec{v}_{i}^{\hspace{1pt} \psi}$, take
\begin{equation} \label{excessvelocity}
\vec{v}_{i}^{\hspace{1pt} \psi} = \vec{v}_{i} - \vec{v}\hspace{1pt}^{\psi}.
\end{equation}
By design, the collection of these excess velocities satisfies a kinematic relation
\begin{equation} \label{Generalconstraint}
\sum_{j=1}^{n} \psi_{j} \vec{v}\hspace{1pt}^{\psi}_{j} = 0.
\end{equation}
In the language of linear algebra, this says that the space of excess species velocities is orthogonal to $\boldsymbol{\psi}.$\footnote{Two subspaces of an inner-product space are said to be orthogonal if the inner product between any two vectors from each of the respective subspaces is zero.} Thus we may also refer to the column $\boldsymbol{\psi}$ itself as the kinematic relation associated with convective velocity $\vec{v}\hspace{1pt}^{\psi}$, because it defines the orthogonal subspace of the species velocities in which the set of excess species velocities must reside.

\subsection{Inversion of the isothermal drag matrix}\label{inversionofM} 

Invariance of the isothermal drag matrix $\textbf{M}$ with respect to the choice of convective velocity requires that
\begin{equation} \label{Lawagainbutwithpsi}
\vec{d}_{i} =  \sum_{j=1}^{n} M_{ij} \vec{v}_{j} =\sum_{j=1}^{n} M_{ij} \vec{v}_{j}^{\hspace{1pt} \psi}
\end{equation}
for any valid kinematic relation $\boldsymbol{\psi}$. Bearing in mind equation \eqref{Generalconstraint}, one can augment the transport matrix and write equation \eqref{Lawagainbutwithpsi} equivalently as 
\begin{equation}
\vec{d}_{i} = \sum_{j=1}^{n} M_{ij} \vec{v}_{j}^{\hspace{1pt}\psi} + \gamma \psi_{i} \sum_{j=1}^{n} \psi_{j} \vec{v}_{j}^{\hspace{1pt}\psi} = \sum_{j=1}^{n} M^{\psi}_{ij} \left( \gamma \right) \vec{v}_{j}^{\hspace{1pt}\psi}
\end{equation}
for any $\gamma >0$. On the right we have defined the \emph{augmented drag matrix} with respect to the kinematic relation $\boldsymbol{\psi}$ as
\begin{equation} \label{Mpsigamdef}
M^{\psi}_{ij} \left( \gamma \right) = M_{ij} + \gamma  \psi_{i} \psi_{j}, ~~\textrm{or}~~ \textbf{M}^\psi \left( \gamma \right) = \textbf{M} + \gamma \boldsymbol{\psi} \boldsymbol{\psi}^\top.
\end{equation}
This augmentation was first considered in the context of multicomponent transport by Helfand \cite{Helfand1960}; much later it was developed independently by Giovangigli \cite{GIOVANGIGLI199073,ERN1995105,giovangigli1999multicomponent}.

Significantly, the augmented transport matrix $\textbf{M}^\psi \left( \gamma \right)$ is positive definite \cite{Vanbruntandother}, and is consequently invertible. Upon inversion one finds that the excess velocity of $i$ is
\begin{equation} \label{Augmentedinversion}
\vec{v}_{i}^{\hspace{1pt}\psi} = \sum_{j=1}^{n} \left[ \textbf{M}^{\psi} \left( \gamma \right) \right]^{-1}_{ij} \vec{d}_{j},
\end{equation}
the desired flux-explicit form of the transport equations. In its present state this formulation is not satisfactory, however. First, the `penalty term' involving $\gamma$ is artificial. Second, and of deeper physical importance, the kinematic relation from equation \eqref{Generalconstraint} is not obviously enforced for every collection of driving forces. 

An unambiguous formulation of inverted transport laws is found by defining
\begin{equation} \label{Lpsibyalimit}
\textbf{L}^{\psi} = \lim_{\gamma \rightarrow \infty} \left[\textbf{M}^{\psi} \left( \gamma \right) \right]^{-1}.
\end{equation}
This limit exists for any valid choice of $\textbf{M}$ and $\boldsymbol{\psi}$. Indeed, 
 \begin{equation}\label{Minvgammaasymptotics}
\left[\textbf{M}^{\psi} \left( \gamma \right) \right]^{-1} = \textbf{L}^{\psi} + \mathscr{O}(\gamma^{-1}),
\end{equation} 
and the unique matrix  $\textbf{L}^\psi$, which we call the \emph{Onsager diffusion matrix} relative to the $\psi$-average velocity, is symmetric whenever $\textbf{M}$ is. This argument is substantiated in \ref{MtoLandback}, which also provides two algebraic formulas useful to compute $\textbf{L}^{\psi}$ from $\textbf{M}$ without resorting to a limit process. 

Since equation \eqref{Augmentedinversion} holds for any nonzero $\gamma$, one can derive the flux-explicit form of the transport laws by calculating the limit of equation \eqref{Augmentedinversion} as $\gamma \to \infty$, giving
\begin{equation} \label{Fluxexplicit}
\vec{v}^{\hspace{1pt}\psi}_{i} = \sum_{j=1}^n L^{\psi}_{ij} \vec{d}_{j}
\end{equation}
as a consequence of the asymptotic behavior in equation \eqref{Minvgammaasymptotics}.  The matrix $\textbf{L}^{\psi}$ so formed also satisfies
\begin{equation} \label{nullspace}
\sum_{j=1}^{n} L^{\psi}_{ij} \psi_{j} = 0,
\end{equation}
that is to say, the kinematic relation $\boldsymbol{\psi}$ is a null eigenvector of $\textbf{L}^{\psi}$. Analogous to the spectral structure of $\textbf{M}$ enforcing the Gibbs--Duhem equation, this spectral structure of $\textbf{L}^{\psi}$ enforces kinematic constraint \eqref{Generalconstraint}.

The procedure outlined above encompasses all methods currently used to invert isothermal Onsager--Stefan--Maxwell equations. Newman's inverse, relative to $\vec{v}_k$, is produced by choosing a kinematic relation such that $\psi_i = \delta_{ik}$, where $\delta_{ij}$ is the Kronecker delta. Given the diffusion driving forces in equation \eqref{didef}, which make $\textbf{1}$ the null eigenvector of $\textbf{M}$, Helfand's inverse is formed by taking $\boldsymbol{\psi} = \textbf{1}$, an unorthodox choice where the arithmetic mean of the species velocities drives convection. Curtiss and Bird's inverse derives from the kinematic relation for the mass-average velocity, $\psi_i = \rho_i$. Crucially, the need for kinematic relations here indicates that one cannot consider a generic matrix $\textbf{L}$ in a flux-explicit thermodiffusion model. Instead there is a class of equally valid Onsager diffusion matrices, from which one is determined by specifying a convective velocity.

Sometimes it may be of interest to calculate $\textbf{M}$ given an $\textbf{L}^{\psi}$, as discussed by Monroe and Newman in a specific case \cite{MONROE20094804}.  This can be accomplished generally by augmenting the singular matrix $\textbf{L}^{\psi}$ to make
\begin{equation}
\textbf{L}^{\psi} \left( \gamma \right) = \textbf{L}^{\psi} + \gamma \textbf{1} \textbf{1}^{\top},
\end{equation}
again with $\gamma > 0$. Then, similar to the process given above, one can pass into a limit to get
\begin{equation}
\textbf{M} = \lim_{\gamma \rightarrow \infty} \left[ \textbf{L}^{\psi} \left( \gamma \right) \right]^{-1}.
\end{equation}
 The $\textbf{M}$ form may be preferable when reporting experimental measurements, since it is independent of the convective velocity used for data processing; the $\textbf{L}^{\psi}$ form, however, provides alternative routes for physical interpretation \cite{Shapiro}.  Nevertheless, one should bear in mind that although there exist many legitimate $\textbf{L}^{\psi}$ matrices, there is just one valid choice of $\textbf{M}$ for a proper set of diffusion driving forces. 

\subsection{Inversion of the non-isothermal drag matrix} \label{subsec:nonisothermalinversion}
Problems posed by the convection invariance of $\textbf{M}$ carry over into the non-isothermal drag matrix $\bigtilde{\textbf{M}}$: the null eigenvector $[ \begin{array}{cc} 0 & \textbf{1}^\top \end{array} ]^{\top}$ induced by convection invariance implies that $\bigtilde{\textbf{M}}$ cannot be directly inverted. It is, however, straightforward to adapt the pseudo-inversion procedure developed earlier to the non-isothermal case.

Thermal properties in the generalized Stefan--Maxwell equations merit some additional discussion here. Observe that only differences between Soret diffusivities appear in equation \eqref{GeneralMSequations}. Since the physical content of that equation is invariant to a variable change where the same scalar is added to every Soret diffusivity, it follows that the collection $\left\{ \mathscr{D}_i^T \right\}_{i=1}^n$ must be linearly dependent. To quantify Soret diffusivities in isolation, one must adopt an additional constraint to close this degree of freedom. Newman circumvented the ambiguity by defining a new property,
\begin{equation} \label{DTintoA}
\mathscr{A}_{ij} = \frac{D_{i}^{T}}{\rho_{i}} - \frac{D_{j}^{T}}{\rho_{j}} = \mathscr{D}_{i}^{T} - \mathscr{D}_{j}^{T},
\end{equation}
which we call the \emph{Newman--Soret diffusivity}. As well as having units $\si{m^2.s^{-1}}$, the fact that $\mathscr{A}_{ii} = 0$ ensures that the nontrivial $\mathscr{A}_{ij}$ are linearly independent for a given $j$. Newman further asserts that these diffusivities are more approximately constant with respect to composition in binary systems \cite{newman1995thermal}. His choice of symbol emphasizes antisymmetry in the indices, that is, that $\mathscr{A}_{ji} = - \mathscr{A}_{ij}$. 

It is standard to enforce a kinematic relation to close the degree of freedom when specifying the coefficients of thermal diffusion. Here we impose the general constraint
 \begin{equation} \label{Thermaldiffusioncoefficientsconstraint}
 \sum_{j=1}^{n} \psi_{j} \mathscr{D}_{j}^T = 0,
 \end{equation}
which is agnostic to the choice of convective velocity. Given any complete set of $\mathscr{A}_{ij}$ with $j$ fixed, the identities $\mathscr{A}_{kl} = \mathscr{A}_{kj}-\mathscr{A}_{lj}$ and $\mathscr{A}_{lk} = - \mathscr{A}_{kl}$ fill out the whole matrix of Newman--Soret diffusivities. Thus, any particular Soret diffusivity can be computed with the formula
\begin{equation} \label{AintoDT}
\mathscr{D}_i^T = \frac{1}{\psi_{\textrm{T}}} \sum_{k=1}^n \psi_k \mathscr{A}_{ik},
\end{equation}
so long as one row or column of the Newman--Soret matrix is known. Given the kinematic relation $\boldsymbol{\psi}$, equations \eqref{DTintoA} and \eqref{AintoDT} establish a bijective map between the Soret diffusivities and the Newman--Soret matrix.

In their implementation of Chapman--Enskog theory, Hirschfelder et al.~state \cite{hirschfelder1954molecular} that 
\begin{equation} \label{DiT_kinematic}
\sum_{i=1}^n D_i^T = 0.
\end{equation}
This implies that the properties $D_i^T / \rho_i$ satisfy a kinematic relation $\psi_i = \rho_i$; the Soret diffusivities stand relative to the mass-average velocity.  Brenner argued that some constitutive laws are more appropriately formulated with respect the volume-average velocity, however \cite{BRENNER2009902}. Rather than advocating one view or the other, we adopt Newman--Soret diffusivities here, which emphasizes that thermal diffusion can be fully parametrized without choosing a reference velocity for convection.

One can write a new equivalent of equation \eqref{CouplednonisothermalM} in terms of the augmented drag matrix $\textbf{M}^\psi \left( \gamma \right)$ from equation \eqref{Mpsigamdef}, as
 \begin{equation} \label{AugmentedCouplednonisothermalM}
\bigtilde{M}_{0i}^\psi \left( \gamma \right) = -\frac{\bigtilde{M}_{00}}{\rho \hat{C}_p T} \sum_{j=1}^{n} M^{\psi}_{ij} \left( \gamma \right) \mathscr{D}_{j}^T.
\end{equation}
These coefficients can be assembled into an augmented column matrix $\widetilde{\textbf{m}}_0^\psi \left( \gamma \right) = [ \bigtilde{M}_{0i}^\psi \left( \gamma \right) ]^{i=1}_n$. The constraint stated in equation \eqref{Thermaldiffusioncoefficientsconstraint} is natural because it implies that
\begin{equation} \label{Augementedm0equalsm0}
\widetilde{\textbf{m}}_0^\psi \left( \gamma \right) = \widetilde{\textbf{m}}_0
\end{equation}
for any $\gamma > 0$. Thus augmentation of the non-isothermal drag matrix has no impact on the thermal entries of $\bigtilde{\textbf{M}}$.

The non-isothermal drag matrix augmented by kinematic relation $\boldsymbol{\psi}$ is defined as
\begin{equation} \label{Mtildepsigamdef}
\bigtilde{\textbf{M}}^{\psi} \left( \gamma \right) =  \begin{bmatrix}
\bigtilde{M}_{00}  & \widetilde{\textbf{m}}_0^\top \\
\widetilde{\textbf{m}}_0 & \textbf{M}^{\psi} \left( \gamma \right) + \frac{ \widetilde{\textbf{m}}_0 \widetilde{\textbf{m}}_0^\top }{\bigtilde{M}_{00}}  
\end{bmatrix},
\end{equation}
in which only the lower-right block depends on the penalty factor $\gamma$. The inverse of this augmented matrix is partially calculable through Schur's complement, further details about which can be found in the book by Zhang \cite{zhang2005schur}.  In block form,
\begin{equation}\label{Schurenough}
\left[ \bigtilde{\textbf{M}}^{\psi} \left( \gamma \right) \right]^{-1} =
  \begin{bmatrix}
\frac{1}{\bigtilde{M}_{00}}+ \frac{\widetilde{\textbf{m}}_{0}^{\top} \left[\textbf{M}^\psi \left( \gamma \right) \right]^{-1} \widetilde{\textbf{m}}_{0}}{\bigtilde{M}_{00}^{2}}     & - \frac{\widetilde{\textbf{m}}_{0}^\top \left[\textbf{M}^\psi \left( \gamma \right) \right]^{-1}}{\bigtilde{M}_{00}}  \\
- \frac{\left[\textbf{M}^\psi \left( \gamma \right) \right]^{-1} \widetilde{\textbf{m}}_{0}}{\bigtilde{M}_{00}}  & \left[\textbf{M}^\psi \left( \gamma \right) \right]^{-1}
\end{bmatrix}.
\end{equation}
Interestingly, the only matrix inversion needed to carry out this computation is that of the augmented isothermal drag matrix.

It is straightforward to rephrase equation \eqref{AugmentedCouplednonisothermalM} in a form that computes Soret diffusivities from the Onsager drag coefficients. The construction
\begin{equation} \label{DiTcolumn}
\left[ \mathscr{D}_{i}^T \right]^{i=1}_n = - \frac{\rho \hat{C}_p T}{\bigtilde{M}_{00} } \left[ \textbf{M}^\psi \left( \gamma \right) \right]^{-1} \widetilde{\textbf{m}}_{0}
\end{equation}
retrieves a column of Soret diffusivities constrained by equation \eqref{Thermaldiffusioncoefficientsconstraint}. With equation \eqref{DiTcolumn}, one can define for $i \in \left\{ 1, ..., n \right\}$ that
\begin{equation} \label{L0idef}
\bigtilde{L}_{0i}^\psi = \frac{\mathscr{D}_i^T}{\rho \hat{C}_p T},
\end{equation}
and introduce a column $\bigtilde{\textbf{l}}^{\hspace{1pt}\psi}_0 = [ \bigtilde{L}_{0i}^\psi ]^{i=1}_n $. (The superscript $\psi$ on these quantities emphasizes kinematic relationship \eqref{Thermaldiffusioncoefficientsconstraint}, i.e. that $\boldsymbol{\psi}^\top \bigtilde{\textbf{l}}^{\hspace{1pt}\psi}_0 = 0$.) Then
\begin{equation} \label{littlelpsidef}
\bigtilde{\textbf{l}}^{\hspace{1pt}\psi}_0 = - \frac{\left[\textbf{M}^\psi \left( \gamma \right) \right]^{-1} \widetilde{\textbf{m}}_{0}}{\bigtilde{M}_{00}}
\end{equation}
for any $\gamma > 0$---indeed, $\bigtilde{\textbf{l}}^{\hspace{1pt}\psi}_0$ is independent of $\gamma$---and the inverse augmented drag matrix transforms to
\begin{equation} \label{reallySchurnow}
\left[ \bigtilde{\textbf{M}}^{\psi} \left( \gamma \right) \right]^{-1} =
  \begin{bmatrix}
\frac{1}{\bigtilde{M}_{00}} \left( 1 - \widetilde{\textbf{m}}_{0}^{\top} \bigtilde{\textbf{l}}^{\hspace{1pt}\psi}_0 \right)     & \left( \bigtilde{\textbf{l}}^{\hspace{1pt}\psi}_0 \right)^\top \\
\bigtilde{\textbf{l}}^{\hspace{1pt}\psi}_0 & \left[\textbf{M}^\psi \left( \gamma \right) \right]^{-1}
\end{bmatrix} .
\end{equation}
Finally, the unique inverse of the non-isothermal drag matrix $\bigtilde{\textbf{M}}$ with respect to convective velocity $\vec{v}^{\hspace{1pt}\psi}$ is found by taking the limit of equation \eqref{reallySchurnow} as $\gamma \to \infty$. This produces the \emph{Onsager thermodiffusion matrix}
\begin{equation} \label{Nonisothermalinverse}
\bigtilde{\textbf{L}}^{\psi} = 
  \begin{bmatrix}
\bigtilde{M}_{00}^{-1} \left( 1 - \widetilde{\textbf{m}}_{0}^{\top} \bigtilde{\textbf{l}}^{\hspace{1pt}\psi}_0 \right)     & \left( \bigtilde{\textbf{l}}^{\hspace{1pt}\psi}_0 \right)^\top \\
\bigtilde{\textbf{l}}^{\hspace{1pt}\psi}_0 & \textbf{L}^\psi
\end{bmatrix} ,
\end{equation}
a symmetric matrix with null eigenvector $[ \begin{array}{cc} 0 & \boldsymbol{\psi}^\top \end{array} ]^\top $. 

\subsection{Onsager--Fick--Fourier laws}
Equation \eqref{Nonisothermalinverse} gives rise to a velocity-explicit form of the non-isothermal flux laws similar to isothermal equation \eqref{Fluxexplicit}. A thermodynamically consistent form of these flux laws is
\begin{equation} \label{Nonisothermalfluxexplicit}
\vec{v}_{i}^{\hspace{1pt}\psi} = \sum_{j=0}^{n} \bigtilde{L}^{\psi}_{ij} \vec{d}_{j},
\end{equation}
where $i$ ranges from $0$ to $n$. Here it should be understood that $\vec{v}_{0}^{\hspace{1pt}\psi} = \vec{v}_{0}$, because the thermal velocity $\vec{v}_0$ is not included in the kinematic constraint from equation \eqref{Generalconstraint}. Unlike the intuitive extension of Fick's law involving transport matrix $\bigtilde{\textbf{D}}$, the derivation of $\bigtilde{\textbf{L}}^{\hspace{1pt}\psi}$ from the Onsager--Stefan--Maxwell thermodiffusion equations ensures that the coefficient matrix in equation \eqref{Nonisothermalfluxexplicit} has consistent units, as well as being symmetric positive semidefinite and satisfying a kinematic constraint. 

Equation \eqref{Nonisothermalfluxexplicit} substantiates a thermodynamically consistent set of generalized Fick--Fourier thermodiffusion laws as follows. After inserting equations \eqref{thermalv}, \eqref{d0ofT}, and \eqref{L0idef} into equation \eqref{Nonisothermalfluxexplicit}, and defining the excess molar flux of species $i$ relative to the $\psi$-average velocity as
\begin{equation} \label{Jipsidef}
\vec{J}_i^{\hspace{1pt}\psi} = c_i \vec{v}_{i}^{\hspace{1pt} \psi},
\end{equation}
one obtains flux-explicit constitutive equations 
\begin{align} \label{genFourierL}
\vec{q}\hspace{1.5pt}' & = - \rho^2 \hat{C}_p^2 T \bigtilde{L}_{00}^{\psi} \vec{\nabla} T + \sum_{j=1}^n \mathscr{D}_{j}^T \vec{d}_j, \\ \label{genFickL}\vec{J}_i^{\hspace{1pt}\psi} & = -c_i \mathscr{D}_i^T \vec{\nabla} \ln T + \sum_{j=1}^n c_i L_{ij}^{\psi}\vec{d}_j,
\end{align}
for $i \in \left\{ 1, ..., n \right\}$. Equation \eqref{genFourierL} is analogous to Fourier's law, with extra terms describing the Dufour effect; equations \eqref{genFickL} generalize Fick's law, with an additional term for the Soret effect. These equations satisfy Onsager's principles of irreversible thermodynamics, in that they involve all the proper fluxes and driving forces that appear in the dissipation functional, as well as including a positive-semidefinite transport matrix that satisfies a symmetric reciprocal relation and has a single null eigenvector to enforce convection invariance. To emphasize their thermodynamic rigour, we henceforth call equations \eqref{genFourierL} and \eqref{genFickL} the \emph{Onsager--Fick--Fourier equations}.

\subsection{Perspectives on thermal conductivity} \label{subsec:thermalconductivity}
To elucidate simple heat conduction, it is worth returning to the dissipation functional \eqref{Fundmentalentropyproduction}. Insertion of definitions \eqref{thermalv} and \eqref{d0ofT} produces the compact expression
\begin{equation} \label{simplestTs}
T \dot{s} = \sum_{i=0}^n \vec{d}_i \cdot \vec{v}_i.
\end{equation}
Putting the Onsager--Stefan--Maxwell thermodiffusion laws \eqref{Forceexplicit2}-\eqref{Forceexplicit1} into equation \eqref{simplestTs} produces equation \eqref{dissipationMbar2}, which shows that the dissipation associated with temperature gradients when the species velocities vanish is proportional to $1/\bigtilde{M}_{00}$. Past analyses have asserted on this basis that $\bigtilde{M}_{00}$ should be inversely proportional to the thermal conductivity \cite{deGrootMazur,Krishna1979}; in that case force-explicit equation \eqref{Forceexplicit2} reduces to Fourier's law of heat conduction in the absence of relative species velocities. 

Alternatively, inserting Onsager--Fick--Fourier equations \eqref{Nonisothermalfluxexplicit} into equation \eqref{simplestTs} and applying Gibbs--Duhem relation \eqref{Gibbs-Duhem} leads to
\begin{equation} \label{dissipationLbar}
T \dot{s} = \sum_{i,j=0}^n \vec{d}_i \cdot \bigtilde{L}^{\psi}_{ij} \vec{d}_{j},
\end{equation}
showing that the dissipation associated with temperature gradients in the absence of diffusion driving forces is proportional to $\bigtilde{L}_{00}$. One could alternatively argue that thermal conductivity should be the coefficient of $- \vec{\nabla} T$ in equation \eqref{genFourierL}, so that it reduces to Fourier's law in the absence of diffusion driving forces. 

Newman pointed out that these two perspectives on thermal conductivity are mutually exclusive \cite{newman1995thermal}. Indeed, after inserting the definitions of $\widetilde{\textbf{m}}_0$ and $\bigtilde{\textbf{l}}_{0}^{\hspace{1pt}\psi}$ into equation \eqref{genFourierL} and applying an identity similar to the one presented in equation \eqref{diffusionaldissipationMandD}, one finds that
\begin{equation} \label{L00ofM00andDs}
\bigtilde{L}_{00}^{\hspace{1pt}\psi} = \frac{1}{\bigtilde{M}_{00}} + \frac{R}{2 \rho^2 \hat{C}_p^2T} \sum_{i,j=1}^n \frac{c_i c_j \mathscr{A}_{ij}^2}{c_\textrm{T} \mathscr{D}_{ij} },
\end{equation}
so the thermal conductivity cannot be proportional to both $1/\bigtilde{M}_{00}$ and $\bigtilde{L}_{00}$ simultaneously. Note that we have included Newman--Soret diffusivities to emphasize that this relationship is independent of the convective velocity, addressing another concern raised by Newman \cite{newman1995thermal}. 

The appropriate definition of thermal conductivity is decided by adopting a standard approach to its experimental measurement. Thermal conductivity in fluids is typically measured by inducing a temperature gradient across a one-dimensional cell with closed ends, enforcing no-molar-flux boundary conditions. When such an apparatus reaches a steady state, the differential material balances demand that species velocities vanish throughout the cell, as well as at its ends. For this experiment, equation \eqref{Forceexplicit2} is appropriate, and it reduces to Fourier's law if
\begin{equation} \label{kdef}
k = \frac{\rho^2 \hat{C}_p^2 T}{\bigtilde{M}_{00}}
\end{equation}
defines the thermal conductivity $k$. The dissipation functional from equation \eqref{dissipationMbar2} requires that $k \ge 0$. 

One consequence of this definition of $k$ is that the generalized Fourier's law contains temperature-gradient-driven heat flux in addition to $- k \vec{\nabla} T$. The extra contribution, which Bird, Stewart, and Lightfoot call the \emph{Dufour flux} \cite{BSL2nd}, can be quantified by inserting equations \eqref{L00ofM00andDs} and \eqref{kdef} into equation \eqref{genFourierL}. One practical problem the experimental definition of $k$ raises is that thermodynamic consistency demands \emph{a priori} knowledge of the Soret diffusivities when applying the Onsager--Fick--Fourier laws. 

Our choice of thermal conductivity also has implications for thermodynamic stability. Since the second law of thermodynamics requires that $k \ge 0$, the condition that dissipation must be positive in equation \eqref{L00ofM00andDs} without diffusion driving forces places a lower bound on $\bigtilde{L}_{00}^{\hspace{1pt}\psi}$, namely
\begin{equation} \label{kstab}
\bigtilde{L}_{00}^{\hspace{1pt}\psi} \ge \frac{R}{2 \rho^2 \hat{C}_p^2T} \sum_{i,j=1}^n \frac{c_i c_j \mathscr{A}_{ij}^2}{c_\textrm{T} \mathscr{D}_{ij} }.
\end{equation}
Here inclusion of the Newman--Soret diffusivity emphasizes this condition's convection invariance. 

\section{Properties and balances for viscous fluids} \label{Balances}

Thermodiffusion is simulated by solving a system of differential balance equations that governs the transient spatial distributions of temperature and composition. As well as necessitating a parametrization of the transport coefficients that make up $\bigtilde{\textbf{M}}$ or $\bigtilde{\textbf{L}}^\psi$, closure of this system requires various equilibrium material properties. Generally these parameters are not constants, but rather state functions dependent on local temperature, pressure, and composition. A key advantage of the irreversible thermodynamics methodology is that every equilibrium state function ultimately derives from the system's free energy. 

\subsection{Diffusion driving forces}
Since molar Gibbs energy depends on temperature, pressure, and composition, all of the material properties depend locally on these basis variables, as well. Thus, elementary thermodynamic dependences imply that each diffusion driving force in a viscous fluid depends on the natural basis variables $T$, $p$, and $\left\{c_i\right\}_{i=1}^n$ as
\begin{equation} \label{realdi}
\vec{d}_i =  - RTc_\textrm{T} \sum_{j=1}^n \chi_{ij} 
\vec{\nabla} \left( \frac{c_j}{c_\textrm{T}} \right) + 
c_i \left( \frac{\overline{m}_i}{\rho} - \overline{V}_i \right) \vec{\nabla} p ,
\end{equation}
where $\overline{V}_i$ is the partial molar volume of species $i$ and $\chi_{ij}$ quantifies how the change in chemical activity of species $i$ varies with the mole fraction of species $j$ (i.e., $c_j/c_\textrm{T}$) at fixed $T$ and $p$.  Notably, the temperature-gradient term in the original definition of $\vec{d}_i$, equation \eqref{didef}, cancels out when chemical-potential gradients are expanded over the natural basis. Terms involving composition gradients in equation \eqref{realdi} express the forces that drive mass diffusion; the remaining term drives pressure diffusion.

Entries within the $n \times n$ activity-gradient matrix $\boldsymbol{\chi}$ must adhere to many constraints. When mixing is thermodynamically ideal, $\boldsymbol{\chi} = \textbf{I}$, where $\textbf{I}$ is the identity matrix. Therefore, one conventionally lets $\boldsymbol{\chi} = \textbf{I} + \boldsymbol{\Gamma}$, such that the $n \times n$ matrix $\boldsymbol{\Gamma}$ represents deviations from ideal mixing behaviour under isothermal, isobaric conditions; Gibbs--Duhem relationships suggest that $\textbf{1}^\top \boldsymbol{\Gamma} = \textbf{o}^\top$. A constraint that $\boldsymbol{\Gamma} \textbf{1} = \textbf{o}$ ensures that nonphysical composition changes lie in the nullspace of $\boldsymbol{\Gamma}$. Finally, although $\boldsymbol{\Gamma}$ is not generally symmetric, Maxwell relations still reduce its number of independent entries to $n \left( n - 1 \right)/2$, one for each distinct pair of species. \ref{nonidealmixing} leverages results of a recent paper by Goyal and Monroe \cite{Goyal2021} to produce a formula for $\boldsymbol{\Gamma}$ in terms of experimentally measurable properties, and \ref{Maxwellrelations} shows how Maxwell relations constrain the entries of $\boldsymbol{\chi}$.

\subsection{Balance equations}

Every multicomponent mass-transport process is governed by a set of species material balances,\footnote{Generally, the material balances can be augmented by generation terms, for example, to account for the formation or consumption of species by homogeneous chemical reactions. Such phenomena, which produce commensurate generation terms in the thermal energy balance as well, are neglected here for simplicity.} 
\begin{equation} \label{substantialmatbal}
\frac{D c_i}{D t} = - \vec{\nabla} \cdot \left( c_i \vec{v}_i^{\hspace{1pt}\rho} \right)- c_i \vec{\nabla} \cdot  \vec{v}\hspace{1pt}^\rho  
\end{equation}
for $i \in \left\{1, ..., n\right\}$. Here the operator
\begin{equation}
\frac{D}{Dt} = \frac{\partial}{\partial t} + \vec{v}\hspace{1pt}^\rho \cdot \vec{\nabla}
\end{equation}
represents the substantial derivative with respect to the mass-average velocity. 

Thermodynamic considerations reveal that the material balances imply a connection between changes of temperature and pressure. A state equation for the volume derives directly from the Euler equation for the Gibbs energy, by differentiation with respect to pressure. The molar volume, $1/c_\textrm{T}$, is an intensive equilibrium property and consequently depends generally on $T$, $p$, and at most $n-1$ species fractions \cite{Goyal01012017}. Thus the local concentrations satisfy
\begin{equation} \label{GibbsPhaseRule}
\sum_{i=1}^n \overline{V}_i c_i = 1,
\end{equation}
in which the partial molar volumes depend on the same basis variables as $c_\textrm{T}$. Since $\phi_i = \overline{V}_i c_i$ represents the volume fraction occupied by species $i$ within a phase, this can be seen as a rather trivial statement, that species volume fractions sum to unity. But it also establishes the Gibbs phase rule, a more substantive physical conclusion.

Equation \eqref{GibbsPhaseRule} leads to the Gibbs phase rule as follows. A substantial derivative of equation \eqref{GibbsPhaseRule} combines with the Gibbs--Duhem relation derived from the extensivity of volume \cite{Goyal01012017} to give
\begin{equation} \label{substantialp}
\frac{Dp}{Dt} = K \alpha_V \frac{DT}{Dt} + K \sum_{i=1}^n \overline{V}_i \frac{Dc_i}{Dt} , 
\end{equation}
where $K$ is the bulk modulus (inverse isothermal compressibility) and $\alpha_V$ is the volumetric coefficient of thermal expansion. Therefore, in an $n$-ary single phase, one of the intensive natural variables $T$, $p$, $\left\{c_i\right\}_{i=1}^n$ necessarily depends on the others.

To close the governing system's degree of freedom associated with temperature, a balance of thermal energy is added alongside the material balances. In any nonreactive single-phase viscous fluid, this heat balance is \cite{Goyal01012017}
\begin{eqnarray} \label{thermalenergybalance}
\rho \hat{C}_p \frac{DT}{Dt} = - \vec{\nabla} \cdot \vec{q}\hspace{1.5pt}' - \sum_{i=1}^n c_i \vec{v}_i^{\hspace{1pt}\rho} \cdot \vec{\nabla} \overline{H}_i + T \alpha_V \frac{Dp}{Dt} \nonumber \\ 
-\ \left(  \vec{\vec{\tau}} + \sum_{i=1}^n c_i \overline{m}_i \vec{v}_i^{\hspace{1pt}\rho} \vec{v}_i^{\hspace{1pt}\rho} \right) : \vec{\nabla} \vec{v}\hspace{1pt}^\rho  ,
\end{eqnarray}
in which $\overline{H}_i$ is the partial molar enthalpy of species $i$ and $\vec{\vec{\tau}}$ is deformation stress, the latter following Bird, Stewart, and Lightfoot's sign convention \cite{BSL2nd}. The terms on the right of equation \eqref{thermalenergybalance} respectively quantify local heat accumulation from: net influx of irreversible heat; excess enthalpy convection due to diffusion; externally imposed compressive power; and viscous friction. 

\subsection{Species enthalpies} 

The heat balance apparently introduces a new set of properties --- the partial molar enthalpies --- but much of the information they contain is embedded in quantities that have already been defined. In light of the standard Legendre transformation $\mu_i = \overline{H}_i - T \overline{S}_i$, the definition of the diffusion driving force and equation \eqref{realdi} give
\begin{equation} \label{firstHi}
\vec{\nabla} \overline{H}_i = \overline{V}_i \vec{\nabla} p + \frac{RTc_\textrm{T}}{c_i} \sum_{j=1}^n \chi_{ij} \vec{\nabla} y_j + T \vec{\nabla}\overline{S}_i.
\end{equation}
Also, $\overline{S}_i$ connects to the chemical potential through a Maxwell relation, so the last term depends on existing parameters. As discussed in \ref{nonidealmixing}, one ultimately finds
\begin{align}
\vec{\nabla} \overline{H}_i = \overline{C}_{p,i} \vec{\nabla} T + \left[ 1 - \left( \frac{\partial \ln \overline{V}_i }{\partial \ln T} \right)_p \right] \overline{V}_i \vec{\nabla} p  \\
-\frac{RTc_\textrm{T}}{c_i} \sum_{j=1}^n \left( \frac{\partial \chi_{ij} }{\partial \ln T} \right)_p \vec{\nabla} \left( \frac{c_j}{c_\textrm{T}} \right) , \label{Hibarreal}
\end{align}
in which $\overline{C}_{p,i}$ represents the partial molar constant-pressure heat capacity of species $i$. 

Substituting material balances \eqref{substantialmatbal} into equation \eqref{substantialp} shows that the time derivative of pressure in equation \eqref{thermalenergybalance} depends only on spatial derivatives and the time derivative of temperature. Insertion of equation \eqref{Hibarreal} then yields an explicit equation for the time evolution of temperature as a function of the instantaneous spatial distributions of temperature, pressure, composition, the set of excess species velocities $\left\{ \vec{v}_i^{\hspace{1pt}\rho} \right\}_{i=1}^n$, and the mass-average velocity $\vec{v}\hspace{1pt}^\rho$. Since the Euler equation for entropy implies that
\begin{equation} \label{rhoCpofbarCpi}
\rho \hat{C}_p = \sum_{i=1}^n c_i \overline{C}_{p,i},
\end{equation}
this heat balance is fully parametrized by specifying the temperature, pressure, and composition dependences of $\alpha_V$, $K$, $\left\{\overline{C}_{p,i}\right\}_{i=1}^n$, $\left\{\overline{V}_{i}\right\}_{i=1}^n$, the molar masses $\left\{ \overline{m}_i \right\}_{i=1}^n$, and the $n ( n-1 ) /2$ independent state functions that underpin $\boldsymbol{\chi}$. Bearing in mind that mass density satisfies
\begin{equation} \label{rhodef}
\rho = \sum_{i=1}^n \overline{m}_i c_i 
\end{equation}
by definition, the parameters that have already been stated suffice to quantify the diffusion driving forces, as well.

\subsection{Dissipation}

The only quantity that remains free in heat balance \eqref{thermalenergybalance} is the deformation stress. In general, the dissipation function for a nonreactive viscous fluid is \cite{Goyal01012017}
\begin{equation}
T \dot{s} = \sum_{i=0}^n \vec{d}_i \cdot \vec{v}_i -\left( \vec{\vec{\tau}} + \sum_{i=1}^n c_i \overline{m}_i \vec{v}_i^{\hspace{1pt}\rho} \vec{v}_i^{\hspace{1pt}\rho} \right) : \vec{\nabla} \vec{v}\hspace{1pt}^\rho,
\end{equation}
which augments equation \eqref{simplestTs} to account for viscosity. Viscous dissipation was neglected up to now to keep the discussion simpler: as a consequence of Curie's principle, which states that transport processes are decoupled if their associated fluxes differ in tensor rank by one \cite{Prigogine}, the addition of viscous dissipation to the functional does not affect the derivation of Onsager--Stefan--Maxwell laws. Closing a model that includes $\vec{\vec{\tau}}$ requires a distinct constitutive law for stress, such as Newton's law of viscosity. 

\section{Thermodiffusion in monatomic-gas mixtures}

Numerous measurements have been made to quantify thermodiffusion in binary noble-gas mixtures. To put this body of experimental data in context, we first lay out equilibrium properties and the thermal balance equation for multicomponent gas mixtures.  We subsequently discuss the physical assumptions that underpin experimental data processing, and tabulate the measured thermodiffusion properties for various noble-gas pairs. 

\subsection{Equilibrium properties}
Here we restrict the development to ideal mixtures of monatomic gases, for which a substantial amount of experimental data is available. The equilibrium material properties take simpler forms in this case, greatly simplifying both the diffusion driving forces and the energy balance. 

As well as satisfying $\boldsymbol{\chi}=\textbf{I}$, an ideal gas is defined in part by the condition that $\overline{V}_i = RT/p$ for all $i$. Thus equation \eqref{GibbsPhaseRule} yields
\begin{equation} \label{pV=nRT}
c_\textrm{T} = \frac{p}{RT},
\end{equation}
a relationship which can be used to demonstrate that $K = p$ and $\alpha_V = 1 / T$, both independent of composition. Also, this form of the partial molar volumes reduces equation \eqref{Hibarreal} to
\begin{equation} \label{Hibarideal}
\vec{\nabla} \overline{H}_i = \overline{C}_{p,i} \vec{\nabla} T.
\end{equation}
Generally the ideal-gas state is also defined by a condition that each   species' partial molar heat capacity $\overline{C}_{p,i}$ in a gas mixture equals the molar heat capacity of the pure species in isolation. (Recall that the molar enthalpy of any pure species in the ideal-gas state depends on temperature alone.)  

In an ideal gas, equation \eqref{realdi} for the force driving diffusion of species $i$ reduces to 
\begin{equation} \label{Drivingforcesideal}
\vec{d}_{i} = -p \nabla y_{i} + y_{i} \left( \frac{\overline{m}_i c_\textrm{T}}{\rho}- 1 \right) \vec{\nabla} p, 
\end{equation}
where we have eliminated $c_\textrm{T}$ with state equation \eqref{pV=nRT} and defined the mole fraction $y_{i} = c_{i}/c_{\text{T}}$ to simplify notation. Thus the driving force for pressure diffusion is proportional to the difference between the molar mass of species $i$ and the number-average molar mass of the mixture, $\rho / c_\textrm{T} = \sum_{i=1}^n \overline{m}_i y_i$.

The equilibrium thermal properties simplify even more if all the species comprising an ideal gas are assumed to be monatomic. In a monatomic ideal gas, $\overline{C}_{p,i} = \frac{5}{2}R$ for each $i$. Through equation \eqref{rhoCpofbarCpi}, the volumetric heat capacity of a monatomic ideal-gas mixture is therefore $\rho \hat{C}_p = \frac{5}{2} c_\textrm{T} R$, independent of composition.

\subsection{Practical constraints}

As it presently stands, the governing system is one vector equation short of closure. Letting $d$ be the number of spatial dimensions, transport is governed by pressure continuity equation \eqref{substantialp}, heat balance \eqref{thermalenergybalance}, $n$ material balances \eqref{substantialmatbal}, $d(n+1)$ Onsager--Fick--Fourier flux-law components \eqref{Nonisothermalfluxexplicit}, and $d^2$ viscous-fluid constitutive-law components. These govern the transient spatial distributions of $T$, $p$, $\left\{ c_i \right\}_{i=1}^n$, $\vec{q}\hspace{1.5pt}'$, $\{ \vec{v}_i^{\hspace{1pt}\rho} \}_{i=1}^n$, $\vec{v}\hspace{1pt}^\rho$, and $\vec{\vec{\tau}}$. Thus the model comprises $\left[ 2+n+d\left( 1+ n + d \right) \right]$ scalar equations in $\left[ 2+n+d\left( 2 + n + d \right) \right]$ scalar unknowns, leaving $d$ degrees of freedom.

 The final degrees of freedom can be closed naturally by including a detailed local momentum balance, such as Cauchy's equation, to specify how the convective velocity evolves in response to the pressure and deformation-stress distributions. This has not been done in practice, however, in part because most thermodiffusion experiments probe steady states, use closed containers, and assume one-dimensional geometries. Thus no-molar-flux conditions at the ends of the container enforce a condition that all species velocities vanish everywhere.

In practice, experimentalists studying gases have simplified models with an isobaric assumption, taking $p$ to be spatially uniform,  but this does not suffice to close the degrees of freedom when transport occurs across more than one spatial dimension.  For isobaric gas thermodiffusion, the condition that $\vec{\nabla} p = \vec{0}$ simplifies diffusion driving forces \eqref{Drivingforcesideal} to 
\begin{equation} \label{Drivingforcesidealisobaric}
\vec{d}_{i} = -p \vec{\nabla} y_{i} .
\end{equation}
The assumed spatial uniformity still allows a time dependence of pressure, however, so the substantial derivatives in equations \eqref{substantialp} and \eqref{thermalenergybalance} reduce to $d p / d t$, rather than vanishing. Thus, although the isobaric assumption simplifies the governing equations, it does not fully resolve the closure problem, because it leaves the balance of equations and unknowns unchanged. 

 In the following discussion,  we will neglect viscous effects for consistency with the thermodiffusion literature, and will assume an isobaric condition with $\vec{v}\hspace{1pt}^\rho = \vec{0}$ in place of a formal momentum balance. This simplification is consistent with literature assumptions for one-dimensional systems. For simulations of closed containers with more spatial dimensions at steady state, it amounts to an additional stipulation that the mass-average velocity is irrotational. 

\subsection{Thermodiffusion property measurements}
Experiments to parametrize thermodiffusion typically employ binary gas mixtures. The measurement is performed in an apparatus designed to be one-dimensional spatially, which is subjected to a temperature difference across its ends. Designating one end of the apparatus by $a$ and the other by $b$, experimentalists record the steady-state value of the parameter
\begin{equation} \label{Thermalseparationexperiment}
\alpha_{12} = \frac{\ln \left( y_{2}^{b}/y_{1}^{b} \right) - \ln \left( y_{2}^{a} / y_{1}^{a} \right) }{ \ln T^{b}/T^{a} }, 
\end{equation}
known as the thermal diffusion factor \cite{Grew1952}. Conventionally, index $1$ is assigned to the heavier of the two species.

An expression relating $\alpha_{12}$ to transport properties derives naturally from the Onsager--Stefan--Maxwell laws. In a steady state, the condition $\vec{v}^{\rho} = 0$ simplifies material balances \eqref{substantialmatbal} dramatically, to 
\begin{equation} \label{SSmatbalvrho0}
0 = -\vec{\nabla} \cdot \left( c_i \vec{v}_i^{\hspace{1pt}\rho} \right )
\end{equation}
for each species $i$. In a one-dimensional apparatus with closed ends, these balances imply that $\vec{v}_i = \vec{0}$ for every species $i$. Thus the force-explicit flux law \eqref{Forceexplicit2} can be used to eliminate $\vec{v}_0$ from equation \eqref{Forceexplicit1}, showing that
\begin{equation} \label{diforexperiment}
\vec{d}_i = \frac{\bigtilde{M}_{i0} }{\bigtilde{M}_{00}} \vec{d}_0
\end{equation}
for each species $i=1, \dots, n$. Substitution of driving forces \eqref{d0ofT} and \eqref{Drivingforcesidealisobaric} into relation \eqref{diforexperiment}, followed by simplification with equation of state \eqref{pV=nRT} and rearrangement based on the fact that $y_2 = 1 - y_1$, give
\begin{equation}
\vec{\nabla} \ln \left( \frac{y_2}{y_1} \right) = \frac{\mathscr{A}_{12}}{\mathscr{D}_{12} } \vec{\nabla} \ln T 
\end{equation}
for an isobaric binary mixture of monatomic gases in a closed, one-dimensional apparatus subjected to a steady temperature polarization. Integration from end $a$ to end $b$ yields 
\begin{equation} \label{equation for thermal separation}
\alpha_{12} = \frac{\mathscr{A}_{12}}{\mathscr{D}_{12}},
\end{equation}
assuming that the ratio $\mathscr{A}_{12}/\mathscr{D}_{12}$ is constant with respect to temperature and composition. Thus the thermal diffusion factor relates directly to the Newman--Soret coefficient and the Stefan--Maxwell diffusivity.

Table \ref{Datatable} presents the binary diffusion coefficients and thermal diffusion factors for various pairs of noble gases. 
\begin{table}
\begin{center}
\begin{tabular}{ p{1.4cm}|p{2.9cm}|p{2.1cm} } 
\toprule
Gas pair & $p \mathscr{D}_{12}$ ($\si{atm. cm^2 s^{-1}}$) & $\alpha_{12}$ (unitless) \\ 
\midrule
\ce{Ne} - \ce{He}   & 1.1079 & 0.3432  \\ 
\ce{Ar} - \ce{He}  & 0.7560   & 0.3984  \\
\ce{Kr} - \ce{He} &0.6553 & 0.4279 \\
\ce{Xe} - \ce{He}   &0.5611 & 0.4250 \\
\ce{Ar} - \ce{Ne} & 0.3250 & 0.1741 \\
\ce{Kr} - \ce{Ne}  & 0.2647 & 0.2710  \\
\ce{Xe} - \ce{Ne} & 0.2231 & 0.2951  \\
\ce{Kr} - \ce{Ar} &0.1398 & 0.0705 \\
\ce{Kr} - \ce{Ar}   &0.1144 & 0.0803  \\
\ce{Xe} - \ce{Kr} & 0.0745 & 0.0262 \\
\bottomrule
\end{tabular}
\end{center}
\caption{Thermodiffusion transport coefficients measured for noble-gas pairs at $T=300~\si{K}$  as reported by Mason and Marrero \cite{Mason1972} and Taylor \cite{Taylor1980}.  In all cases data were gathered from  experimental  systems which comprise uniform equimolar mixtures when relaxed to isothermal equilibrium.}
\label{Datatable}
\end{table}
In the first-order approximation of the Chapman--Enskog expansion, the terms $p \mathscr{D}_{12}$ and $\alpha_{12}$ are independent of pressure \cite{hirschfelder1954molecular}.  Hence the reported values are nominally valid for a range of pressures.
Binary diffusion coefficients were computed with the least-squares equations compiled in the comprehensive review by  Mason and Marrero \cite{Mason1972}. The thermal separation factors were taken from temperature-dependent least-squares correlations reported by Taylor \cite{Taylor1980}. 

Many additional measurements of thermal gas-mixture separation exist in the literature. The reported data cover hydrogen/noble-gas mixtures \cite{Dunlop1987}, mixtures of hydrogen isotopes \cite{Grew1966}, halogen/noble-gas mixtures \cite{Maya1974Br2, Edelstein1972}, and gas mixtures that include methane \cite{DUNLOP1987584}.

Complete specification of the thermodiffusion transport matrix requires a value for the thermal conductivity. Thermal conductivities of pure noble gases are reported in Table \ref{ThermalconductivityDatatable}, taken from the survey of Kestin et al.~\cite{Kestin1984}. Note that all gases are understood to be at their average isotopic compositions.
\begin{table}
\begin{center}
\begin{tabular}{ p{1.5cm}|p{2.6cm} } 
\toprule
Gas  & $k$ ($\si{W.cm^{-1}.K^{-1}}$) \\
\midrule
\ce{He} &  15.66   \\ 
\ce{Ne} & 4.98 \\ 
\ce{Ar}  &1.78  \\
\ce{Kr} & 0.95 \\
\ce{Xe} & 0.55 \\

\bottomrule
\end{tabular}
\end{center}
\caption{Thermal conductivities $k$ for pure noble gases at $300~\si{K}$  at atmospheric pressure and average isotopic composition \cite{Kestin1984}. }
\label{ThermalconductivityDatatable}
\end{table}
Even simulating binary systems leads to complications because the thermal conductivity is a strong function of composition; for example, the thermal conductivities of xenon and helium differ by well over an order of magnitude. Thus it is necessary to incorporate a mixing rule for thermal conductivities, a topic for which there is a significant body of literature. Options include the Wassijewa model \cite{Wassiljewa1904} based on the kinetic theory of gases, the empirical Kennard mixing rule \cite{Kennard1938}, and the Mason--Saxena mixing rule \cite{Mason1958}. Recently, an accurate and easily implemented mixing rule based on equivalent-circuit analysis was proposed by Udoetok \cite{Etim2013}.

\section{Numerical computation}
\label{sec:Numerical_computation}

Practical modelling of thermodiffusion has been impeded by the difficulty of numerical implementation. On one hand, deriving Onsager--Fick--Fourier laws for multicomponent mixtures is a complex procedure, and, as we have illustrated, the parametrization of the Onsager thermodiffusion matrix $\bigtilde{\textbf{L}}^\psi$ depends strongly on the choice of reference velocity. On the other hand, in the generalized Stefan--Maxwell equations \eqref{GeneralMSequations}, the mixture of fluxes and driving forces apparently yields no useful mathematical structure to facilitate numerical algorithms. 

Recently the authors proposed an effective finite element discretization for the isothermal Onsager--Stefan--Maxwell equations \cite{Vanbruntandother}, formulated with an arbitrary number of species. Simulations were designed to solve the force-explicit transport equations for gases directly, coupling them to mass-continuity equations for each species. Although that work used the mass-average velocity as convective reference, it is straightforward to make alternative choices. For further analysis of the finite-element solver, we direct the reader to reference \cite{Vanbruntandother}.

\subsection{Extension of binary data}

The Onsager--Stefan--Maxwell equations allow a natural extrapolation of isothermal property data from binary systems to multicomponent systems. Reasoning physically that species/species drag is dominated by pairwise interactions to a first approximation, one can use  diffusion coefficients $\mathscr{D}_{ij}$ measured from binary systems directly in simulations with more than two species. A similar process cannot be used to infer the set of Newman--Soret diffusivities $\mathscr{A}_{ij}$ from binary measurements, however, as a consequence of the structure required by equation \eqref{DTintoA}. Producing a consistent form of the multicomponent Newman--Soret matrix requires that the binary data be regularized to ensure kinematic consistency. \ref{sec:construction_soret} presents a regularization process underpinned by the method of least squares.

\subsection{Steady thermodiffusion in a ternary noble-gas mixture}

The Onsager--Stefan--Maxwell formalism casts thermodiffusion in such a way that heat can be handled as another species within the transport constitutive laws, allowing ready extension of the force-explicit solver developed previously \cite{Vanbruntandother}. In the extended solver, flux laws \eqref{Forceexplicit2} and \eqref{Forceexplicit1} are coupled not only to $n$ material balances in the form of equation \eqref{substantialmatbal}, but also to a thermal energy balance like equation \eqref{thermalenergybalance}.

Here we consider steady, three-dimensional thermodiffusion in an isobaric ternary mixture of noble gases. In this situation, flux laws \eqref{Forceexplicit2}-\eqref{Forceexplicit1} for indices $i \in \left\{ 0, 1, 2, 3 \right\}$ are written in compact form as
\begin{equation}
\vec{d}_i = \sum_{j=0}^3 \bigtilde{M}_{ij} \vec{v}_j,
\end{equation}
where it is understood that $\vec{d}_0 = - \frac{5}{2} c_\textrm{T} R \vec{\nabla} T$ and  $\vec{v}_0 = \vec{q}\hspace{1.5pt}' / ( \tfrac{5}{2} c_\textrm{T} R T ) $. To close the model we adopt the condition $\vec{v}^{\rho} = 0$, which simplifies the material balances to the form of equation \eqref{SSmatbalvrho0} for each $i \in \left\{ 1, 2, 3 \right\}$. For a monatomic-gas mixture, the steady-state heat balance simplifies to
\begin{equation}
0 = -  \frac{5}{2} \vec{\nabla}  \cdot ( c_\textrm{T} R T \vec{v}_0 ),
\end{equation}
and equation \eqref{Drivingforcesideal} describes the diffusion driving forces.

\subsection{Example}
\begin{figure}
\centering
\includegraphics[width=3.25in]{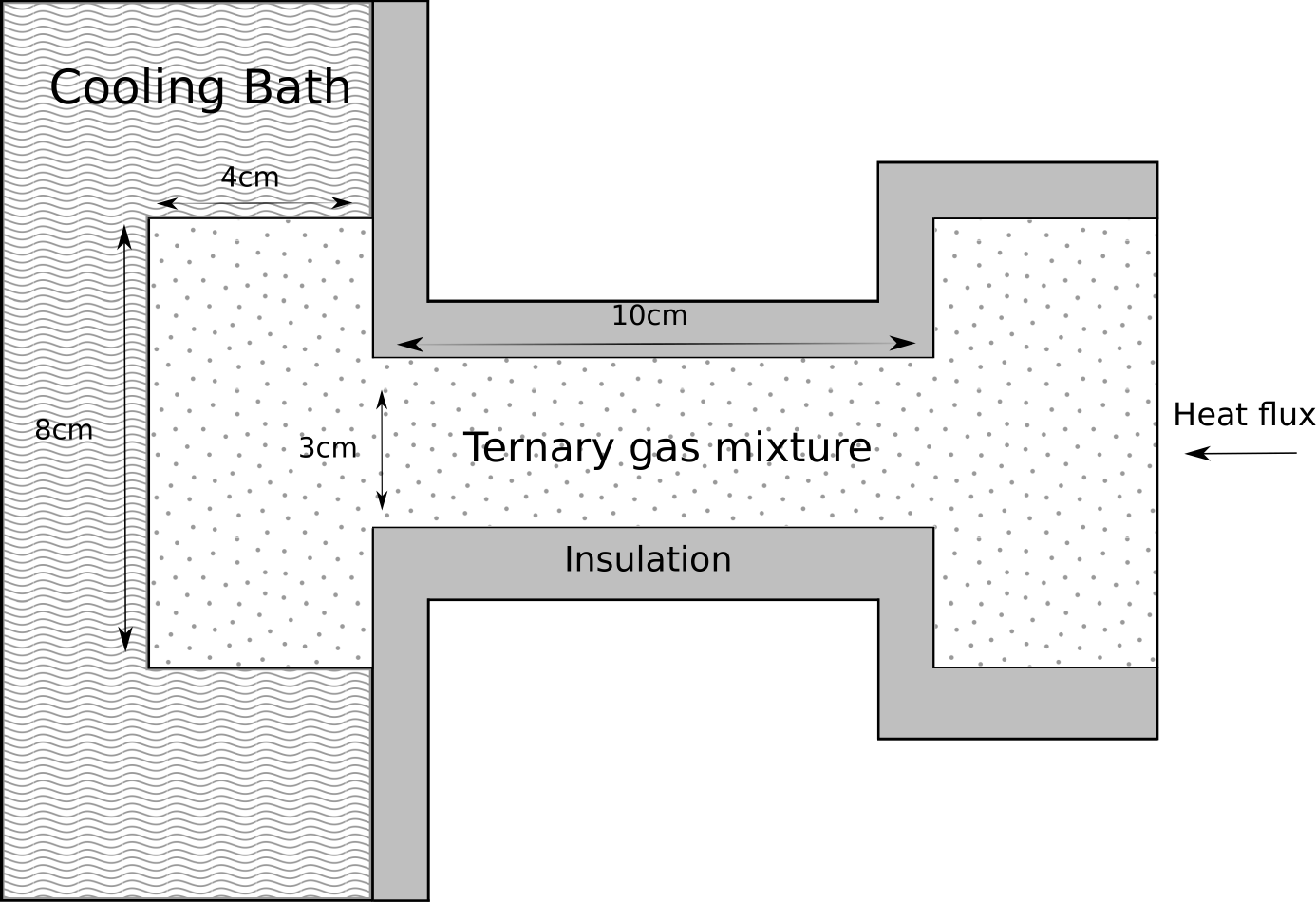}
\caption{Schematic diagram of the separation chamber. The figure shows a cross section of the device, which is symmetric with respect to rotation about its central axis.}
\label{Separation schematic}
\end{figure}

We illustrate the numerical method by simulating the separation of an equimolar ternary gas mixture of helium, argon and krypton in a three-dimensional separation chamber.  The conceptual schematic for this numerical experiment is shown in Figure \ref{Separation schematic}.

Values of binary Stefan--Maxwell coefficients $\mathscr{D}_{ij}$ were taken from Table \ref{Datatable}. Newman--Soret coefficients $\mathscr{A}_{ij}$ computed using binary data for $\alpha_{12}$ are tabulated in Table \ref{BinaryAijtable}, and the regularized Newman--Soret matrix used for simulations is presented in Table \ref{Aijtable}. Thermal conductivities for each pure component were taken from Table \ref{ThermalconductivityDatatable}; to evaluate the thermal conductivity of the mixture, the mixing rule proposed by Udoetok \cite{Etim2013} was employed. This mixing rule was chosen due to its good balance between accuracy and simplicity, the latter of which is key because conductivity must be computed dynamically from the local composition and temperature at each point in the simulation domain. The exact molar masses used for the simulation were $4.00 \: \si{g.mol^{-1}}$, $39.95 \: \si{g.mol^{-1}}$, and $83.80 \: \si{g.mol^{-1}}$ for helium, argon, and krypton, respectively.

The simulation was performed with Firedrake software \cite{Rathgeber2016}, using the MUMPS direct linear solver \cite{MUMPS01, MUMPS02} via PETSc \cite{petsc-user-ref, petsc-efficient}. The mesh of the geometry was constructed using the Gmsh software \cite{GMSH}. Each linear system had 3,895,568 degrees of freedom and convergence was achieved in 11 non-linear iterations. The exact scripts used to produce each numerical experiment can be found at \url{https://bitbucket.org/vanbrunt/consolidated-thermodiffusion-repository} along with the mesh used for the separation chamber.  The exact software versions used to produce the results in this paper, along with instructions for installation, have been archived on Zenodo \cite{zenodo/Firedrake-20211106.0}.

\begin{table}
\begin{center}
\begin{tabular}{ p{1.5cm}|p{1.5cm} p{1.5cm} p{1.5cm}  } 
\toprule
\multicolumn{4}{c}{ $\mathscr{A}_{ij}$ from binary measurements} \\
\midrule
  & \ce{He} \: & \ce{Ar} \: & \ce{Kr} \: \\ 
\midrule

\ce{He} & $0$ & $-0.3012$ & $-0.2804$  \\
\ce{Ar}   & $0.3012$ & $0$ & $-0.0099$  \\
\ce{Kr}  & $0.2804$ & $0.0099$  & $0$ \\
\bottomrule
\end{tabular}
\end{center}
\caption{Newman--Soret diffusivities  computed using the experimental data from binary systems summarized in Table \ref{Datatable} and equation \eqref{equation for thermal separation}. }
\label{BinaryAijtable}
\end{table}

\begin{table}
\begin{center}
\begin{tabular}{ p{1.5cm}|p{1.5cm} p{1.5cm} p{1.5cm}  } 
\toprule
\multicolumn{4}{c}{ Inferred $\mathscr{A}_{ij}$ entries} \\
\midrule
  & \ce{He} \: & \ce{Ar} \: & \ce{Kr} \: \\ 
\midrule

\ce{He} & $0$ & $-0.2910$ & $-0.2906$  \\
\ce{Ar}   & $0.2910$ & $0$ & $0.004$  \\
\ce{Kr}  & $0.2906$ & $-0.004$  & $0$ \\
\bottomrule
\end{tabular}
\end{center}
\caption{Table of regularized Newman--Soret diffusivities for a He--Ar--Kr mixture, determined from the data in Table \ref{BinaryAijtable} with the procedure detailed in \ref{sec:construction_soret}.}
\label{Aijtable}
\end{table}

For the experiment, the cooling bath was maintained at 300 $\si{K}$, a condition reflected in the simulation by imposing Dirichlet boundary conditions specifying the wall temperatures of the left chamber. To drive the temperature gradient, an evenly-distributed fixed heat flux was fed normally into the rightmost face of the device. The remaining boundary conditions were set as homogeneous Neumann conditions to reflect the presence of insulation and gas-impermeable walls.  Last, it was assumed that the total molar amounts of each gas are constants of the closed system. The constant total amount of material in the separation chamber, $n_\textrm{T}$, necessarily satisfies
\begin{equation} \label{molesconserved}
\int_V c_\textrm{T} dV = n_\textrm{T},
\end{equation}
where $V$ indicates the chamber volume. Through equation of state \eqref{pV=nRT}, this implies that
\begin{equation}
p = \frac{R n_\textrm{T}}{\int_{V} 1/T dV},
\end{equation}
which was used to determine a self-consistent steady-state pressure from the temperature distribution. For simulations, $n_\textrm{T}$ was chosen to be consistent with an average total concentration given by equation \eqref{pV=nRT} with $p =1 \, \si{atm}$ and $T = 300\, \si{K}$.

\begin{figure}
\centering
\includegraphics[width=3.25in]{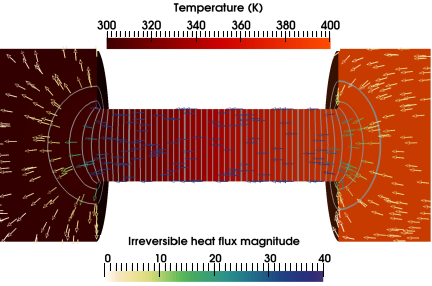}
\caption{Temperature profile within the separation chamber. Arrows indicate the irreversible-heat-flux (\si{W.cm^{-2}}) vector field.}
 \label{Temperaturegradient}
\end{figure}

\begin{figure}
\centering
\includegraphics[width=3.25in]{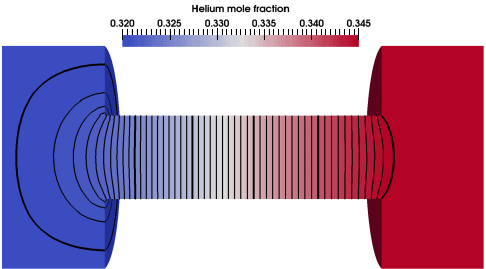}
\caption{Mole fraction profile of helium.}
\label{Heliummolefraction}
\end{figure}

\begin{figure}
\centering
\includegraphics[width=3.25in]{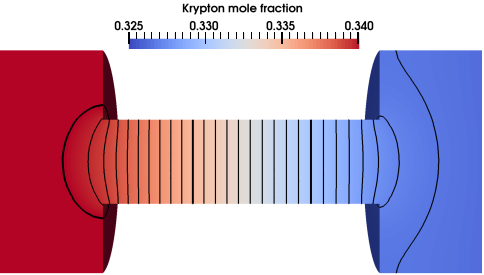}
\caption{Mole fraction profile of krypton.}
\label{Kryptonfraction}
\end{figure}

A heat input of $135\,\si{W}$ induces a temperature difference of approximately $100\, \si{K}$ across the separation chamber, as can be seen from Figure \ref{Temperaturegradient} alongside arrows that indicate the direction and magnitude of local irreversible heat flux.
The separation of helium effected by the temperature gradient is plotted in Figure \ref{Heliummolefraction}. As expected, helium, the lightest component, tends to build up towards the hotter end \cite{Kjelstrup}; a $7.5 \%$ separation of helium is induced by the heat flux.
The steady-state polarization of krypton induced by the heat flux is shown in Figure \ref{Kryptonfraction}. Krypton tends to move towards the cooler end, but the separation is only about $4.5 \%$. 

This numerical example demonstrates the potential of the Onsager--Stefan--Maxwell framework for modelling thermodiffusion in multispecies phases with three-dimensional geometries. From this perspective, one can investigate a rich variety of phenomena for a number of practical applications. To the best of the authors' knowledge, no other multidimensional simulations of thermodiffusion have been performed using the linear irreversible thermodynamics approach.

\section{Conclusion}

A unified framework for modelling fluid thermodiffusion was presented. The new force-explicit Onsager--Stefan--Maxwell formulation was shown to be compatible with the generalized Stefan--Maxwell equations implied by kinetic theory, as well as the commonly employed flux-explicit Onsager--Fick--Fourier constitutive laws. A general method for inverting isothermal transport laws to obtain velocity-explicit forms from force-explicit forms, and vice versa, was proposed, and shown to encompass the disparate prior approaches to flux-law inversion proposed by Newman, Helfand, and Curtiss and Bird. The general inversion process was further extended to cover non-isothermal situations, proving equivalence of the various constitutive formulations.

The Onsager--Stefan--Maxwell thermodiffusion equations involve a non-isothermal transport matrix that proves to have a simple spectral structure. It was shown how to infer entries of multicomponent transport matrices given binary measurements, allowing for straightforward practical implementation of the theory. Further mathematical relationships necessary to implement a thermodiffusion model, such as balance equations and expressions of the driving forces in terms of natural thermodynamic basis variables, were provided for thermodynamically non-ideal viscous fluids and ideal gases. Finally, an effective and novel numerical method to simulate thermodiffusion was deployed, in which heat was treated as a pseudospecies. Multidimensional simulations of steady thermodiffusion in a ternary noble-gas mixture demonstrated the effectiveness of the proposed approach.

Although the numerical methodology was shown to be sound, the absence of convection (i.e., a zero mass-average velocity) is generally not a valid physical assumption. Addressing this issue requires consideration of a local momentum balance alongside the balances of material and heat. Future work is needed to confront the full dynamics of convective thermodiffusion from the irreversible-thermodynamics perspective.

\section*{Acknowledgements} 

This work was supported by the Engineering and Physical Sciences Research Council Centre for Doctoral Training in Partial Differential Equations: Analysis and Applications (EP/L015811/1), the Engineering and Physical Sciences Research Council (grants EP/R029423/1 and EP/V001493/1); the Clarendon fund scholarship; and the Faraday Institution Multiscale Modelling project (subaward FIRG003 under grant EP/P003532/1).

\appendix

\section{Nomenclature}
\setlength{\tabcolsep}{0pt}
\noindent \emph{Greek symbols}
\begin{longtable}{p{0.7cm}p{15.8cm}}
$\alpha_{12}$ & Thermal diffusion factor for separation of species $1$ and $2$ defined in equation \eqref{Thermalseparationexperiment}, [unitless]. By convention $1$ refers to the species with higher $\overline{m}_i$.\\
$\alpha_V$ & Volumetric coefficient of thermal expansion, [\si{K^{-1}}]. \\
$\gamma$ & Penalty factor in the augmented drag matrix $\textbf{M}^\psi \left( \gamma \right)$, see equation \eqref{Mpsigamdef}, [various units]. \\
$\boldsymbol{\Gamma}$ & Matrix representing deviation of $\boldsymbol{\chi}$ from ideality, defined $\boldsymbol{\Gamma} = \boldsymbol{\chi}-\textbf{I}$, $n \times n$ dimensional, [unitless]. \\
$\mu_i$ & Chemical potential of species $i \in \left\{ 1, ..., n \right\}$, [\si{J.mol^{-1}}]. \\ 
$\rho$ & Total mass density, [\si{mol.m^{-3}}]. Expressed in terms of composition by equation \eqref{rhodef}. \\
$\rho_i$ & Mass density of species $i \in \left\{ 1, ..., n \right\}$, [\si{mol.m^{-3}}]. Relates to $c_i$ and $\overline{m}_i$ through $\rho_i = c_i \overline{m}_i$. \\
$\vec{\vec{\tau}}$ & Deformation-stress tensor, [\si{J.m^{-3}}]. \\
$\chi_{ij}$ & Thermodynamic factor quantifying how the change in chemical activity of species $i$ varies with the mole fraction of species $j$, [unitless]. See equation \eqref{realdi}. \\
$\boldsymbol{\chi}$ & Matrix of thermodynamic factors $\chi_{ij}$, $n \times n$ dimensional, [unitless]. \\
$\psi_i$ & Coefficient used when defining a convective velocity, [various units]. See equation \eqref{vpsidef}. \\
$\psi_\textrm{T}$ & Total of coefficients used when defining a convective velocity, [various units]. See equation \eqref{psiTdef}. \\
$\boldsymbol{\psi}$ & Column matrix, $n \times 1$ dimensional, comprising the coefficients $\psi_i$, $\boldsymbol{\psi}= \left[ \psi_i \right]^{i=1}_n$, [various units]. Represents the kinematic relation that constrains excess species velocities $\vec{v}_{i}^{\hspace{1pt} \psi}$. 
\end{longtable}
\noindent \emph{Other symbols}
\begin{longtable}{p{0.7cm}p{15.8cm}}
$\textbf{1}$ & Column matrix whose every entry is `1', $n \times 1$ dimensional, [unitless]. \\
$^\top$ & Matrix transpose operation.
\end{longtable}
\noindent \emph{Roman symbols}
\begin{longtable}{p{0.7cm}p{15.8cm}}
$\mathscr{A}_{ij}$ & Newman--Soret diffusivity, describing thermodiffusion of species $i$ relative to that of species $j$, defined by equation \eqref{DTintoA}, [\si{m^2.s^{-1}}]. \\
$c_i$ & Molar concentration of species $i \in \left\{ 1, ..., n \right\}$, [\si{mol.m^{-3}}]. Relates to $y_i$ and $c_\textrm{T}$ through $c_i = y_i c_\textrm{T}$. \\
$c_\textrm{T}$ & Total molarity, defined $c_\textrm{T} = \sum_{i=1}^n c_i$, [\si{mol.m^{-3}}]. Constrained thermodynamically by equation of state \eqref{GibbsPhaseRule}. \\
$\hat{C}_p$ & Specific constant-pressure heat capacity, [\si{J.K^{-1}.kg^{-1}}]. Relates to $\overline{C}_{p,i}$ through equation \eqref{rhoCpofbarCpi}.  \\
$\overline{C}_{p,i}$ & Partial molar constant-pressure heat capacity of species $i \in \left\{ 1, ..., n \right\}$, [\si{J.K^{-1}.mol^{-1}}]. Relates to $\hat{C}_p$ through equation \eqref{rhoCpofbarCpi}. \\
$\vec{d}_{i}$ & Force driving thermodiffusion of quantity $i$, [\si{J.m^{-4}}]. Equation \eqref{d0ofT} defines the thermal driving force $\vec{d}_0$; equation \eqref{didef} defines the force driving diffusion for every species $i \in \left\{ 1, ...., n \right\}$. \\
$D_{i}^T$ & Coefficient of thermal diffusion for species $i$, [\si{J.s.m^{-3}}]. 
Relates to $\mathscr{D}_{i}^T$ through $D_i^T = \rho_i \mathscr{D}_{i}^T$. \\
$\mathscr{D}_{i}^T$ & Soret diffusivity of species $i$, [\si{m^{2}.s^{-1}}]. Relates to $D_{i}^T$ through equation \eqref{Soretdiffusivitydef}. \\
$\mathscr{D}_{ij}$ & Stefan--Maxwell diffusivity of species $i \in \left\{ 1, ...., n \right\}$ through species $j \in \left\{ 1, ...., n \right\}$, [\si{m^{2}.s^{-1}}]. Only defined when $i \ne j$. Relates to $\textbf{M}$ through equation \eqref{transport matrix}. \\
$\overline{H}_i$ & Partial molar enthalpy of species $i \in \left\{ 1, ..., n \right\}$ appearing in equations \eqref{firstHi} and \eqref{Hibarreal}, [\si{J.mol^{-1}}]. Satisfies Legendre transformation $\mu_i = \overline{H}_i - T\overline{S}_i$. \\
$\textbf{I}$ & Identity matrix, [unitless]. \\
$\vec{J}_i^{\hspace{1pt}\psi}$ & Vector describing the excess molar flux of species $i \in \left\{ 1, ...., n \right\}$ relative to the $\psi$-average velocity defined in equation \eqref{Jipsidef}, [\si{mol.m^{-2}.s^{-1}}]. \\
$k$ & Thermal conductivity, defined by equation \eqref{kdef}, [\si{J.K^{-1}.m^{-1}.s^{-1}}]. \\
$K$ & Bulk modulus (or inverse isothermal compressibility), [\si{J.m^{-3}}]. \\
$\bigtilde{\textbf{l}}^{\hspace{1pt}\psi}_0$ & Column matrix, $n \times 1$ dimensional, made up of the Onsager diffusivities that map diffusion driving forces into Soret fluxes relative to $\boldsymbol{\psi}$, defined in equation \eqref{littlelpsidef}, [\si{m^{5}.J^{-1}.s^{-1}}]. \\
$L^{\psi}_{ij}$ & Onsager diffusion coefficient of species $i$ through $j$ relative to the $\psi$-average velocity, [\si{m^{5}.J^{-1}.s^{-1}}]. \\
$\textbf{L}^\psi$ & Onsager diffusion matrix relative to $\boldsymbol{\psi}$, defined in equation \eqref{Minvgammaasymptotics}, [\si{m^{5}.J^{-1}.s^{-1}}]. \\
$ \bigtilde{\textbf{L}}^{\psi} $ & Onsager thermodiffusion matrix relative to $\boldsymbol{\psi}$, defined in equation \eqref{Nonisothermalinverse}, [\si{m^{5}.J^{-1}.s^{-1}}]. \\
$\widetilde{\textbf{m}}_{0}$ & Column matrix, $n \times 1$ dimensional, made up of the Onsager drag coefficients that map species velocities into the Dufour flux. Defined as $\widetilde{\textbf{m}}_{0} = \left[ \bigtilde{M}_{0i} \right]^{i=1}_n$, [\si{J.s.m^{-5}}]. See equation \eqref{Blocknonisotherm} \emph{et seq.} \\
$\widetilde{\textbf{m}}_{0}^\psi$ & Column matrix, $n \times 1$ dimensional, made up of Onsager drag coefficients that map species velocities into the Dufour flux and augmented by $\boldsymbol{\psi}$, [\si{J.s.m^{-5}}]. Entries defined in equation \eqref{AugmentedCouplednonisothermalM}; equals $\widetilde{\textbf{m}}_{0}$ through equation \eqref{Augementedm0equalsm0}. \\
$\overline{m}_i$ & Molar mass of species $i \in \left\{ 1, ..., n \right\}$, [\si{kg.mol^{-1}}]. \\
$M_{ij}$ & Onsager drag coefficient parametrizing the map from species velocities into diffusion driving forces under isothermal conditions, [\si{J.s.m^{-5}}]. \\
$\textbf{M}$ & Isothermal transport matrix, $n \times n$ dimensional 
with entries $M_{ij}$, [\si{J.s.m^{-5}}]. \\
$M^{\psi}_{ij}$ & Entry in the augmented drag matrix with respect to $\boldsymbol{\psi}$ defined in equation \eqref{Mpsigamdef}, [\si{J.s.m^{-5}}]. \\
$\textbf{M}^\psi$ & Augmented drag matrix with respect to $\boldsymbol{\psi}$ defined in equation \eqref{Mpsigamdef}, $n \times n$ dimensional, [\si{J.s.m^{-5}}].  \\
$\bigtilde{M}_{ij}$ & Onsager drag coefficient parametrizing the map from thermodiffusive velocities into thermodiffusion driving forces, [\si{J.s.m^{-5}}]. \\
$\bigtilde{\textbf{M}}$ & Non-isothermal drag matrix, $(n+1)\times(n+1)$ dimensional and positive semidefinite with entries $\bigtilde{M}_{ij}$, [\si{J.s.m^{-5}}]. \\
$\bigtilde{\textbf{M}}^{\psi}$ & Non-isothermal drag matrix augmented by $\boldsymbol{\psi}$, $(n+1)\times(n+1)$ dimensional and positive semidefinite with entries $\bigtilde{M}_{ij}$, [\si{J.s.m^{-5}}]. Defined in equation \eqref{Mtildepsigamdef}. \\
$n$ & Number of species comprising a diffusion or thermodiffusion medium, [unitless]. \\ 
$n_\textrm{T}$ & Total number of moles contained in a thermodiffusion system, [\si{mol}]. \\ 
$\vec{N}_{S}$ & Entropy flux vector, [\si{J.K^{-1}.m^{-2}.s^{-1}}]. \\
$p$ & Absolute external pressure, [\si{J.m^{-3}}]. \\
$\vec{q}\hspace{1.5pt}'$ & Irreversible heat flux vector, [\si{J.m^{-2}.s^{-1}}]. \\
$R$ & Gas constant, \SI{8.31446}{J.K^{-1}.mol^{-1}}. \\
$\dot{s}$ & Volumetric rate of entropy production, [\si{J.K^{-1}.m^{-3}.s^{-1}}]. The rate of energy dissipation is $T \dot{s}$, [\si{J.m^{-3}.s^{-1}}]. \\
$\hat{S}$ & Specific entropy, [\si{J.K^{-1}.kg^{-1}}]. \\
$\overline{S}_i$ & Partial molar entropy of species $i \in \left\{ 1, ..., n \right\}$, [\si{J.K^{-1}.mol^{-1}}]. \\
$t$ & Time, [\si{s}]. \\
$T$ & Absolute temperature, [\si{K}]. \\
$\vec{v}_{i}$ & Thermodiffusive velocity vector of quantity $i$, [\si{m.s^{-1}}]. Equation \eqref{thermalv} defines the thermal velocity $\vec{v}_0$; the species velocities, $\vec{v}_i$ such that $i \in \left\{ 1, ...., n \right\}$, appear first in equation \eqref{Fundmentalentropyproduction}. \\
$\vec{v}_{i}^{\hspace{1pt} \psi}$ & Vector quantifying the velocity of species $i$ relative to the $\psi$-average velocity, defined by equation \eqref{excessvelocity}, [\si{m.s^{-1}}]. \\
$\vec{v}^\rho$ & Mass-average velocity vector, defined by equation \eqref{vrhodef}, [\si{m.s^{-1}}]. \\
$\vec{v}\hspace{1pt}^{\psi}$ & $\psi$-average velocity vector, defined by equation \eqref{vpsidef}, [\si{m.s^{-1}}]. Describes an arbitrary convective velocity. \\
$\vec{v}^c$ & Number-average velocity vector, defined by equation \eqref{vcdef}, [\si{m.s^{-1}}]. \\
$V$ & Chamber volume, see equation \eqref{molesconserved}, [\si{m^3}]. \\
$\overline{V}_i$ & Partial molar volume of species $i \in \left\{ 1, ..., n \right\}$, [\si{m^{3}.mol^{-1}}]. \\
$ y_i $ & Mole fraction of species $i \in \left\{ 1, ..., n \right\}$, defined $y_i = c_i / c_\textrm{T}$, [unitless]. 
\end{longtable}

\section{Inversion of transport matrices}\label{MtoLandback}

Both the procedure that maps $\textbf{M}$ to $\textbf{L}^{\psi}$ for a given kinematic relation $\boldsymbol{\psi}$, and the inverse process, by which $\textbf{L}^{\psi}$ maps to $\textbf{M}$ given the Gibbs--Duhem relation $\textbf{1}$, require some further justification. Here we present formulas to construct $\textbf{L}^{\psi}$ from $\textbf{M}$ and $\textbf{M}$ from $\textbf{L}^{\psi}$, as well as demonstrating that these two mappings are inverses of each other.

Given a symmetric matrix $\textbf{R}$ with a single null eigenvector $\textbf{r}$, and an additional column matrix $\textbf{p}$ such that $\textbf{p}^\top \textbf{r} \ne 0$, the augmented matrix $\textbf{R}^p \left( \gamma \right)$, defined as
\begin{equation} \label{Rpextension}
\textbf{R}^p \left( \gamma \right) = \textbf{R} + \gamma \textbf{p} \textbf{p}^\top,
\end{equation}
has rank one higher than $\textbf{R}$ whenever $\gamma \ne 0$. This construction is amenable to one of the generalized Sherman--Morrison formulas derived by Baksalary et al.~\cite{BAKSALARY2003207}. We employ the notation $\textbf{A}^{\sharp}$ to represent the Moore--Penrose pseudoinverse of $\textbf{A}$. Further note that since there is just one null eigenvector of $\textbf{R}$, the matrix Baksalary and colleagues call $\textbf{Q}_{\textbf{R}^\top}$, which represents the orthogonal projector onto the orthogonal complement of the column space of $\textbf{R}$, can be formed explicitly:
\begin{equation}
\textbf{Q}_{\textbf{R}^\top} 
= \frac{\textbf{r}\textbf{r}^\top}{\textbf{r}^\top\textbf{r}}.
\end{equation}
Equation (2.5) from reference \cite{BAKSALARY2003207} then yields
\begin{equation} \label{Rpsharp}
\left[ \textbf{R}^p \left( \gamma \right) \right]^\sharp = \left( \textbf{I} - \frac{\textbf{r} \textbf{p}^\top}{\textbf{r}^\top \textbf{p}} \right) \textbf{R}^\sharp \left( \textbf{I} - \frac{\textbf{p} \textbf{r}^\top}{\textbf{r}^\top \textbf{p}} \right) + \frac{1}{\gamma} \frac{\textbf{r} \textbf{r}^\top}{( \textbf{r}^\top \textbf{p})^2}
\end{equation}
after some algebraic rearrangement.

Next define the matrix $\textbf{P}$ such that 
\begin{equation}\label{PMoorePenroseform}
\textbf{P} =  \left( \textbf{I} - \frac{\textbf{r} \textbf{p}^\top}{\textbf{r}^\top \textbf{p}} \right) \textbf{R}^\sharp \left( \textbf{I} - \frac{\textbf{p} \textbf{r}^\top}{\textbf{r}^\top \textbf{p}} \right).
\end{equation}
Direct calculation shows that $\textbf{P}$ affords $\textbf{p}$ as a null eigenvector, explaining its notation. The matrix $\textbf{P}$ so formed is unique because $\textbf{R}^\sharp$ is, and the fact that symmetry of $\textbf{R}$ implies symmetry of $\textbf{R}^\sharp$ means $\textbf{P} = \textbf{P}^\top$ as well.

Since $\textbf{R}$ has a single null eigenvector and $\textbf{R}^p \left( \gamma \right)$ has rank one higher than $\textbf{R}$, it follows that $\textbf{R}^p \left( \gamma \right)$ is invertible. Therefore the pseudoinverse $\left[ \textbf{R}^p \left( \gamma \right) \right]^\sharp$ simplifies to the standard matrix inverse; equation \eqref{Rpsharp} rearranges to give
\begin{equation}\label{Pinverseform}
\textbf{P} = \left[ \textbf{R}^p \left( \gamma \right) \right]^{-1} - \frac{1}{\gamma} \frac{\textbf{r} \textbf{r}^\top}{( \textbf{r}^\top \textbf{p})^2}.
\end{equation}
This formula for $\textbf{P}$ is independent of the choice of nonzero $\gamma$ and circumvents the computation of $\textbf{R}^\sharp$. A trivial rearrangement of this result,
\begin{equation} \label{RpofP}
\left[ \textbf{R}^p \left( \gamma \right) \right]^{-1} = \textbf{P} + \frac{1}{\gamma} \frac{\textbf{r} \textbf{r}^\top}{( \textbf{r}^\top \textbf{p})^2},
\end{equation}
substantiates the asymptotic behaviour asserted in equation \eqref{Minvgammaasymptotics}. Direct calculation then shows that
\begin{equation} \label{Plimitform}
\textbf{P} = \lim_{\gamma \to \infty} \left[ \textbf{R}^p \left( \gamma \right) \right]^{-1},
\end{equation}
justifying the claim that the limit in equation \eqref{Lpsibyalimit} exists.

Finally we consider uniqueness of the mapping between $\textbf{P}$ and $\textbf{R}$. Suppose we have a new matrix, $\hat{\textbf{P}}$, whose sole null eigenvector is also the previously defined column matrix $\textbf{p}$, and let $\textbf{r}$ continue to be the sole null eigenvector of $\textbf{R}$. For any $\phi \ne 0$ the augmented matrix
\begin{equation} \label{Profomega}
\hat{\textbf{P}}^r \left( \phi \right) = \hat{\textbf{P}} + \phi \textbf{r} \textbf{r}^\top
\end{equation}
is invertible with rank one higher than $\hat{\textbf{P}}$. As before, the formula of Baksalary et al.~\cite{BAKSALARY2003207} constructs a unique matrix 
\begin{equation}
\hat{\textbf{R}} = \left[ \hat{\textbf{P}}^r \left( \phi \right) \right]^{-1} - \frac{1}{\phi} \frac{\textbf{p}\textbf{p}^\top}{\left( \textbf{r}^\top \textbf{p} \right)^2}.
\end{equation}
Letting $\gamma = \phi^{-1} \left( \textbf{r}^\top \textbf{p} \right)^{-2}$ in equation \eqref{Rpextension} and using that result to eliminate the second term on the right produces an equality which proves that
$\hat{\textbf{R}} = \textbf{R}$ if and only if 
\begin{equation}
\left[ \hat{\textbf{P}}^r \left( \phi \right) \right]^{-1} = \textbf{R}^p \left( \frac{1}{\phi \left( \textbf{r}^\top \textbf{p} \right)^2 } \right).
\end{equation}
Since the matrices here are nonsingular, one can invert both sides, then insert equations \eqref{RpofP} and \eqref{Profomega} to show that
$\hat{\textbf{P}} = \textbf{P}$. Thus our flux-law inversion establishes a bijective mapping between the space of matrices $\textbf{P} = \textbf{P}^\top$ whose nullspaces are spanned by $\textbf{p}$ and the space of matrices $\textbf{R} = \textbf{R}^\top$ whose nullspaces are spanned by $\textbf{r}$.

The transport-law inversions developed in section \ref{inversionofM} can be implemented directly using the general equations provided here. Specifically, to get flux-explicit isothermal transport laws from the Onsager--Stefan--Maxwell formulation, take $\textbf{r}=\textbf{1}$, $\textbf{p} = \boldsymbol{\psi}$, and $\textbf{R} = \textbf{M}$, then use equation \eqref{PMoorePenroseform}, \eqref{Pinverseform}, or \eqref{Plimitform} to produce $\textbf{P} = \textbf{L}^\psi$. Alternatively, to retrieve force-explicit isothermal transport laws from the isothermal Onsager--Fick--Fourier formulation, let $\textbf{r}=\boldsymbol{\psi}$, $\textbf{p} = \textbf{1}$, and $\textbf{R} = \textbf{L}^\psi$, in which case any of the same equations will produce $\textbf{P} = \textbf{M}$.

\section{Non-ideal Darken factors}\label{nonidealmixing}
This appendix focuses on connecting measurable thermodynamic factors to the properties $\chi_{ij}$ in equation \eqref{realdi}. 
Goyal and Monroe analyzed nonideal thermodynamic factors, emphasizing the practical constraint that only $n-1$ composition descriptors can be varied independently, and that species activity coefficients are constrained by Gibbs--Duhem relations \cite{Goyal2021}. They developed several results in terms of Darken factors \cite{Darken48} over the basis of mole fractions $\left\{ y_i \right\}_{i=1}^n$, where $y_i = c_i / c_\textrm{T}$. This choice of basis makes clear the intrinsic composition constraint, because
\begin{equation} \label{molefractionsum}
\sum_{i=1}^n y_i = 1
\end{equation}
everywhere, independent of $T$ and $p$. The Darken factor $\overline{Q}_{ij}$ is defined through the partial derivative
\begin{equation} \label{Qdef}
\overline{Q}_{ij} = y_i \left( \frac{\partial \mu_i}{\partial y_j} \right)_{T,p,y_{k \ne j, n} } .
\end{equation}
If $y_j$ varies, then mole-fraction sum \eqref{molefractionsum} requires at least one other mole fraction to vary. By requiring that the particular mole fraction $y_n$ be left free when taking any other composition derivative, Darken factors defined in this way both ensure compatibility with the mole-fraction sum and clarify what conditions to enforce during characterization experiments. Note that  $\overline{Q}_{ij} = RT \delta_{ij}$ if mixing is ideal. 

This discussion requires matrices with various sizes and shapes, so it will help to notate matrix dimensions. Here, a bold capital letter with a single subscript index will imply a square matrix, with the index indicating the number of rows; a bold lower-case letter with a single subscript index will imply a column matrix, again with the index denoting total rows; row matrices will always be annotated as transposes of columns; and rectangular matrices with $m>1$ rows and $n>1$ columns will be annotated with capital letters and subscript indices $m \times n$. 

Our task here is to relate the nonideal activity-gradient matrix $\boldsymbol{\Gamma}_n$ introduced in section \ref{Balances} to the smaller Darken matrix $\overline{\textbf{Q}}_{n-1}$. As we shall see shortly, this can be done explicitly through matrix operations that exploit an invertible, symmetric matrix $\textbf{G}_{n}$ with constant entries
\begin{equation} \label{GDextension}
\left( \textbf{G}_n \right)_{ij} = \delta_{ij} - \delta_{in}-\delta_{nj}.
\end{equation}
One can write the inverse of $\textbf{G}_n$ as
\begin{equation}
\left( \textbf{G}_n^{-1} \right)_{ij} = \delta_{ij} - \delta_{in} \delta_{nj} - \frac{1}{n} .
\end{equation}
The truncated matrix $\textbf{G}_{n \times (n-1)}$ formed by striking the $n^\textrm{th}$ column of $\textbf{G}_{n}$ represents the linear transformation process Goyal and Monroe call Gibbs--Duhem extension \cite{Goyal2021}. 

In terms of the Darken matrix $\overline{\textbf{Q}}_{n-1}$, the first $n-1$ diffusion driving forces are given by the matrix equation
\begin{equation} \label{dinminus1}
\left[ \vec{d}_i \right]^{i=1}_{n-1} = -c_\textrm{T} \overline{\textbf{Q}}_{n-1} \left[ \vec{\nabla} y_i \right]^{i=1}_{n-1} + \left[ \frac{\rho_i}{\rho} - \phi_i \right]^{i=1}_{n-1} \vec{\nabla} p,
\end{equation}
an expression that only involves the first $n-1$ mole-fraction gradients. Alternatively, all $n$ driving forces are written through equation \eqref{realdi} in a form that includes all $n$ mole-fraction gradients. Multiplication by $\textbf{G}_{n \times (n-1)}$ can extend the column of driving forces in equation \eqref{dinminus1} as
\begin{equation}
\left[ \vec{d}_i \right]^{i=1}_{n} = \textbf{G}_{n \times (n-1)} \left[ \vec{d}_i \right]^{i=1}_{n-1},
\end{equation}
or contract the mole-fraction gradients in equation \eqref{realdi}, through a similar expression where $\vec{\nabla} y_i$ replaces $\vec{d}_i$. Setting equations \eqref{dinminus1} and \eqref{realdi} equal therefore shows that
\begin{equation}\label{Gammarankincomplete}
\boldsymbol{\Gamma}_n \textbf{G}_{n \times (n-1)} = \textbf{G}_{n \times (n-1)} \left( \frac{\overline{\textbf{Q}}_{n-1}}{RT} - \textbf{I}_{n-1} \right) ,
\end{equation}
where the nonideal correction $\boldsymbol{\Gamma}_n$ has been used instead of the thermodynamic matrix $\boldsymbol{\chi}_n$. This specifies most of $\boldsymbol{\Gamma}_n$, but leaves a number of its entries undetermined.

Observe that taking the gradient of the mole-fraction sum and casting the result in matrix form shows that
\begin{equation}
\textbf{1}_n^\top \left[ \vec{\nabla} y_i \right]^{i=1}_{n-1} = \vec{0},
\end{equation}
so the domain of the mapping represented by $\boldsymbol{\Gamma}_n$ lies in a subspace orthogonal to $\textbf{1}_n$. To make $\boldsymbol{\Gamma}_n$ unambiguous, we stipulate that any quantity outside its domain lies in its nullspace, that is, that $\boldsymbol{\Gamma}_n \textbf{1}_n = \textbf{o}_n$. Since direct calculation with equation \eqref{GDextension} shows that $\textbf{1}_n$ is also orthogonal to the column set of $\textbf{G}_{n \times (n-1)}$, one can concisely represent all of the properties of $\boldsymbol{\Gamma}_n$ by augmenting equation \eqref{Gammarankincomplete}:
\begin{equation}
\boldsymbol{\Gamma}_n \textbf{G}_{n} = \left[  \begin{array}{cc} \textbf{G}_{n \times (n-1)} \left( \frac{\overline{\textbf{Q}}_{n-1}}{RT} - \textbf{I}_{n-1}\right)  & \textbf{o}_n \end{array} \right] .
\end{equation}
Right multiplication of both sides by $\textbf{G}_{n}^{-1}$ then shows that
\begin{equation} \label{GammaofQ}
\boldsymbol{\Gamma}_n = \left[  \begin{array}{cc} \textbf{G}_{n \times (n-1)} \left( \frac{\overline{\textbf{Q}}_{n-1}}{RT} - \textbf{I}_{n-1}\right)  & \textbf{o}_n \end{array} \right] \textbf{G}_{n}^{-1},
\end{equation} 
the desired explicit equation for $\boldsymbol{\Gamma}_n$ in terms of $\overline{\textbf{Q}}_{n-1}$. 

Equation \eqref{GammaofQ} is useful when expressing the entropy gradients from equation \eqref{firstHi} in terms of the natural basis variables. Through a Maxwell relation, the composition derivatives of $\overline{S}_i$ follow from equation \eqref{Qdef}:
\begin{equation}
\left( \frac{\partial \overline{S}_i}{\partial y_j} \right)_{T,p,y_{k \ne j, n}} 
= - \frac{1}{y_i} \left( \frac{\partial \overline{Q}_{ij}}{\partial T} \right)_{p,y_k}.
\end{equation}
Recalling that $\boldsymbol{\Gamma}_n = \boldsymbol{\chi}_n - \textbf{I}_n$, Gibbs--Duhem extension can then be used to find
\begin{equation}
\sum_{j=1}^{n-1} \left( \frac{\partial \overline{S}_i}{\partial y_j} \right)_{T,p,y_{k \ne j, n}} \vec{\nabla} y_j = - \frac{R}{y_i} \sum_{j=1}^n \frac{\partial \left( T \chi_{ij} \right)}{\partial T} \vec{\nabla} y_j ,
\end{equation} 
a result that was used when deriving equation \eqref{Hibarreal}.

\section{Maxwell relations for activities} \label{Maxwellrelations}
It is not immediately obvious, but the thermodynamic matrix $\boldsymbol{\chi}_n$ from equation \eqref{realdi} is also constrained by Maxwell relations. (Here we employ the conventions for notating matrix size laid out in \ref{nonidealmixing}.) These constraints are clarified by relating the thermodynamic factors to entries $\overline{K}_{ij}$ within the isothermal, isobaric Hessian of molar Gibbs energy $\overline{G}$ with respect to the independent mole fractions, defined as
\begin{equation}
\overline{K}_{ij} = \left( \frac{\partial^2 \overline{G}}{\partial y_i \partial y_j} \right)_{T,p,y_{k \ne i,j, n}}.
\end{equation}
Thermodynamic stability demands that the Hessian matrix $\overline{\textbf{K}}_{n-1}$ is positive semidefinite \cite{Goyal2021}, and Maxwell relations imply its symmetry \cite{Monroe2015}. 

Monroe and Newman \cite{MONROE20094804} introduced a composition matrix $\textbf{Y}_{n-1}$ with entries 
\begin{equation}
Y_{ij} = \frac{\delta_{ij} }{y_i} + \frac{1}{y_n},
\end{equation}
and showed that it transforms the free-energy Hessian into the Darken matrix (cf.~\ref{nonidealmixing}), as
\begin{equation} \label{QYK}
\overline{\textbf{Q}}_{n-1} = \textbf{Y}_{n-1} \overline{\textbf{K}}_{n-1}.
\end{equation}
It has also been shown that $\textbf{Y}_{n-1}$ is invertible if and only if all of the mole fractions are nonzero \cite{Goyal2021}, so one can isolate $\overline{\textbf{K}}_{n-1}$ here in any situation where all the species are present. In any case, equation \eqref{QYK} establishes that the $(n-1)^2$ entries of $\overline{\textbf{Q}}_{n-1}$ depend explicitly on composition and any set of $n \left( n-1 \right)/2$ independent state functions that specifies the symmetric Hessian $\overline{\textbf{K}}_{n-1}$.

Noting that $\boldsymbol{\Gamma}_n = \boldsymbol{\chi}_n - \textbf{I}_n$, one can substitute equation \eqref{QYK} into equation \eqref{GammaofQ} to get
\begin{equation} \label{chiofK}
\boldsymbol{\chi}_n = \textbf{I}_n + \left[  \begin{array}{cc} \frac{\textbf{G}_{n \times (n-1)} \textbf{Y}_{n-1} \overline{\textbf{K}}_{n-1}}{RT} - \textbf{G}_{n \times (n-1)}  & \textbf{o}_n \end{array} \right] \textbf{G}_{n}^{-1}.
\end{equation} 
Here the symmetry of $\overline{\textbf{K}}_{n-1}$ justifies the claim that all $n^2$ entries of the thermodynamic matrix in equation \eqref{realdi} are parametrized by $n (n-1)/2$ state functions.

\section{Construction of Soret diffusivities} \label{sec:construction_soret}

For $i,j =1,2,...,n$, take as given a set of Newman--Soret diffusivities $ \tilde{\mathscr{A}}_{ij}$ that satisfy $\tilde{\mathscr{A}}_{ij} = - \tilde{\mathscr{A}}_{ij}$ and $ \tilde{\mathscr{A}}_{ii} = 0$. Practically, these would be taken from measurements of each binary system, analogous to the construction of Table \ref{BinaryAijtable}. Consistency of the modelling framework demands that a proper Newman--Soret matrix with entries $\mathscr{A}_{ij}$ be based on a set of Soret diffusivities $\mathscr{D}_{i}$. What constrains us from immediately identifying $\mathscr{A}_{ij}$ with $ \tilde{\mathscr{A}}_{ij}$ is the fact that it is not clear that this extrapolated $\mathscr{A}_{ij}$ has the form
demanded by equation \eqref{DTintoA} for some set of Soret diffusivities $\mathscr{D}_{i}^T$. For example, in a ternary mixture, equation \eqref{DTintoA} implies that $\mathscr{A}_{12}+\mathscr{A}_{23} = \mathscr{A}_{13}$, but inspection of the entries tabulated in Table \ref{BinaryAijtable} shows that this does not hold. Thus the data in $ \tilde{\mathscr{A}}_{ij}$ require some regularization to produce $\mathscr{A}_{ij}$.

To regularize $\mathscr{A}_{ij}$, we choose its entries such that each is as close to possible to $ \tilde{\mathscr{A}}_{ij}$. That is, we impose structure \eqref{DTintoA} on the coefficients, while minimizing the changes, according to the minimization problem
\begin{equation}
\min_{\mathscr{D}_{i}^T \in \mathbb{R}} \sum_{i,j}^{n} \left[ \left( \mathscr{D}_{i}^T - \mathscr{D}_{j}^T \right) -  \tilde{\mathscr{A}}_{ij} \right]^{2},
\end{equation}
subject to the constraint given by equation \eqref{Thermaldiffusioncoefficientsconstraint} with $\boldsymbol{\psi} = \boldsymbol{1}$. A direct computation shows that this minimum is obtained by setting 
\begin{equation}
\mathscr{D}_{i}^T = \frac{1}{n} \sum_{j=1}^{n}  \tilde{\mathscr{A}}_{ij}.
\end{equation}

Note that the kinematic relation chosen for the minimization problem implicitly sets a somewhat nonphysical reference velocity for Soret diffusivities. Since $\mathscr{A}_{ij}$ is independent of the convective reference velocity, however, this mathematically convenient choice is immaterial. To enforce a constraint on $\mathscr{D}_{i}^T$ with another $\boldsymbol{\psi}$ in \eqref{Thermaldiffusioncoefficientsconstraint}, one could simply perform a shift analogous to that stated in equation \eqref{excessvelocity} post facto. 

\bibliography{Thermodiffusion.bib}

\end{document}